\documentclass[traditabstract]{aa} 
\usepackage{graphicx}
\usepackage{lscape}
\usepackage{natbib}
\usepackage{txfonts}
\begin{document}
   \title{X~Her and TX~Psc: Two cases of  ISM interaction with stellar winds  observed by \textit{Herschel}\thanks{{\it Herschel} is an ESA space observatory with science instruments
provided by European-led Principal Investigator consortia and with important participation from NASA. 
}}

   \author{A. Jorissen
          \inst{1}
          \and A. Mayer
          \inst{2}
          \and S. Van Eck
          \inst{1}
          \and R. Ottensamer
          \inst{2,8}
          \and F. Kerschbaum
          \inst{2}
          \and T. Ueta
          \inst{3}
         \and P. Bergman
          \inst{4}
          \and J.A.D.L. Blommaert
          \inst{5}
          \and L. Decin
          \inst{5}
          \and M.A.T. Groenewegen
          \inst{6}
          \and J. Hron
          \inst{2}
          \and W. Nowotny
          \inst{2}
          \and H. Olofsson
          \inst{4}
          \and Th. Posch
          \inst{2}
          \and L.O. Sjouwerman
          \inst{7}
          \and B. Vandenbussche
          \inst{5}
          \and  C. Waelkens
          \inst{5}
          }

\institute{Institut d'Astronomie et d'Astrophysique, Universit\'e Libre de Bruxelles, CP. 226, 
  Boulevard du Triomphe, 1050 Bruxelles, Belgium \\
              \email{svaneck,ajorisse@astro.ulb.ac.be}
\and
University of Vienna, Department of Astronomy, T\"urkenschanzstra\ss e 17, 1180 Wien, Austria
\and
Dept. of Physics and Astronomy, University of Denver, Mail Stop 6900, Denver, CO 80208, USA
\and
Onsala Space Observatory, SE-43992 Onsala, Sweden and Stockholm Observatory, AlbaNova University Centre, SE-10691 Stockholm, Sweden
\and
Instituut voor Sterrenkunde, K.U. Leuven, Celestijnenlaan, 200D, B-3001 Leuven, Belgium  
\and
Koninklijke Sterrenwacht van Belgi\"e, Ringlaan 3, 1180 Brussels, Belgium
\and
National Radio Astronomy Observatory, P.O. Box 0, Socorro, NM 87801, USA
\and
Institute of Computer Vision and Graphics, TU Graz, Infeldgasse 16/II, A-8010 Graz, Austria
 }

   \date{Received ...; accepted ...}

  \abstract
   {The asymptotic giant branch (AGB) stars X~Her and TX~Psc have been imaged at 70 and 160 $\mu$m with the PACS instrument onboard the {\it Herschel} satellite, as part of the large {\it MESS} (Mass loss of Evolved StarS) Guaranteed Time Key Program. The images 
reveal an axisymmetric extended structure with its axis oriented along the space motion of the stars. This extended structure  is very likely to be shaped by  the interaction of the 
wind ejected by the AGB star with the surrounding interstellar medium (ISM). As predicted by numerical simulations, the detailed structure of the wind-ISM interface depends upon the 
relative velocity between star+wind  and the ISM, which is large for these two stars (108 and 55~km~s$^{-1}$ for X~Her and TX~Psc, respectively). In both cases, there is a compact blob upstream whose origin is not fully elucidated, but that could be the signature of some instability in the wind-ISM shock. Deconvolved images of X~Her and TX~Psc reveal several discrete structures along the outermost filaments, which could be Kelvin-Helmholtz vortices. Finally, TX~Psc is surrounded by an almost circular  ring (the signature of the termination shock?) that contrasts with the outer, more structured filaments.  A similar inner circular structure seems to be present in X Her as well, albeit less clearly.}
 
\authorrunning{Jorissen et al.}
\titlerunning{Herschel observations of X Her and TX Psc}
   \maketitle
%

\section{Introduction}

The {\it MESS} (Mass loss of Evolved StarS) Guaranteed Time Key Program \citep{Groenewegen_2011} observed with the PACS and SPIRE instruments onboard the {\it Herschel Space Observatory}  a representative sample of evolved stars, including 78 asymptotic giant branch (AGB) stars and post-AGB stars with the aim of studying their circumstellar environments and mass-loss history by taking advantage of the good angular resolution offered by {\it Herschel} (with a point-spread function of $5\farcs7$ full-width at half maximum  at 70~$\mu$m).

A large variety of shapes (spherical, elliptical, detached, and axisymmetric shells) has been encountered by this program. A similar conclusion is reached for smaller angular scales (1-$10\arcsec$) by the Plateau de Bure/Pico Veleta CO survey \citep{2007ASPC..378..199C,Castro-Carrizo2010}. In addition, Mid-IR and optical surveys \citep{1999ApJS..122..221M, 2000ApJ...528..861U} have shown that the shells exhibit at least axisymmetry by the end of the AGB phase. AGB stars with detached shells in the MESS sample have already been discussed by \citet{2010A&A...518L.140K,Kerschbaum_galagb}, and a synoptic paper is in preparation (Cox et al.). 
The present paper focuses on two spectacular cases, namely the O-rich star X~Her and the C-rich star TX~Psc, which show axisymmetric shells. 

Extended asymmetric structures around AGB stars have been seen in different wavelength bands (which thus probe different regions of the circumstellar environment): in the IR by \citet{1993ApJS...86..517Y}, \citet{2006ApJ...648L..39U,  2008PASJ...60S.407U, 2010A&A...514A..16U}, \citet{2008ApJ...687L..33U}, \citet{2010AAS...21543114G}, \citet{2010A&A...518L.141L}, 	and \citet{2011A&A...528A..29I}, in H~I at 21~cm by \citet{2006MNRAS.365..245G}, \citet{2007AJ....133.2291M}, \citet{2008ApJ...684..603M}, \citet{2010A&A...515A.112L}, and \citet{Matthews2011}, 
in the CO radio lines by \citet{1989A&A...218L...5H}, \citet{1996A&A...310..952K}, \citet{2010A&A...515A.112L}, and \citet{Castro-Carrizo2010}, and in the UV by \citet{2007Natur.448..780M}. These asymmetries can be triggered by several  causes:
\begin{itemize}
\item[-] Wind ejection in a binary system will shape a spiral stream, according to hydrodynamical simulations \citep{1993MNRAS.265..946T, 1996MNRAS.280.1264T, 1999ApJ...523..357M, 2009MNRAS.396.1805H}. The clearest illustration thereof has been observed in the proto-planetary nebula AFGL 3068 \citep{2006A&A...452..257M, 2006IAUS..234..469M}.
\item[-] Asymmetric mass loss in single stars, as predicted by dust-driven wind models for red giants with cool 
spots \citep{Reimers} and 3D models of carbon stars \citep{2008A&A...483..571F}, may produce an off-centred nebula. Such an asymmetric mass loss may also impart a kick on the forming white dwarf, and evidence thereof has been reported in globular clusters \citep{2007MNRAS.381L..70H, 2007MNRAS.382..915H, 2008MNRAS.385..231H, 2008MNRAS.383L..20D, 2009ApJ...695L..20F}.
\item[-] \citet{2000ApJ...545..965M} demonstrated  that a stellar dipole magnetic field can focus an initially isotropic wind toward the equatorial plane, thus producing an equatorial overdensity, 
but \citet{2005AIPC..804...89S} pointed out that in cases where the magnetic field plays a 
global role in the shaping, a binary companion is necessary to 
maintain the field.
\item[-] Interaction between the stellar wind and the interstellar medium (ISM), which is the topic of the present paper.
\end{itemize}

For X~Her and TX~Psc, there are strong indications that   the interaction of their wind with the ISM shapes their axisymmetric shell. First evidence of this kind of interaction have been seen  in planetary nebulae, in the form of 
axisymmetric morphologies oriented along the space motion of the central star \citep[e.g.,][and references therein]{1981ApJ...244..903J, 1985ApJ...288..622R, 1987A&A...181..373K, 1990ApJ...360..173B, 1998ApJ...495..337D}.
Among non-AGB stars, far-IR parabolic arcs attributable to  bow shocks have been discovered around $\alpha$ Ori from IRAS data by \citet{1988ESASP.281a.249S}, and later confirmed by AKARI data \citep{2008PASJ...60S.407U}. Runaway OB stars are another class of stars around
which bow shocks are frequent \citep{1979ApJ...230..782G,1988ApJ...329L..93V,1993ASPC...35..315V,1995AJ....110.2914V}.

\citet{2006MNRAS.365..245G}, \citet{2006ApJ...648L..39U}, and \citet{2007Natur.448..780M}  mentioned the possibility of wind -- ISM interaction  in relation to AGB mass loss, although the wind velocity is very small ($<$~20~km~s$^{-1}$) in comparison with PNe or OB stars ($\sim$~1000~km~s$^{-1}$) and shock structures seemed therefore unlikely.

In this scenario, the AGB wind is slowed down as it sweeps up ISM material. The swept-up material forms a density enhancement that continues to expand because of the internal pressure.
The thermal emission of the dust  at
the shock interface between the stellar wind and the ISM has been detected in the far IR \citep{2006ApJ...648L..39U, 2009ASPC..418..463U}, or directly in the X/UV, as for the bow shock associated with the wind of $o$~Ceti and detected by GALEX \citep{2007Natur.448..780M}.
Images obtained with Spitzer 
and AKARI have confirmed such an interaction scenario for the targets R~Hya, R~Cas, and \emph{o}~Cet \citep{2006ApJ...648L..39U,2008ApJ...687L..33U,2010A&A...514A..16U}. In RX~Lep,
\citet{2008A&A...491..789L} found an H~I  tail extending 0.5~pc southward, as already suggested for Mira, RS~Cnc, and other sources detected in H~I \citep{2006MNRAS.365..245G,2007AJ....133.2291M,2007Natur.448..780M,2008ApJ...684..603M,2010A&A...515A.112L}. 
\citet{2010ApJ...711L..53S} presented ultraviolet GALEX  images of IRC+10~216 (CW~Leo) revealing for the first 
time its bow shock, which is also visible in {\it Herschel} PACS and SPIRE infrared images \citep{2010A&A...518L.141L}.

\begin{figure}[t]
\centering
   \includegraphics[width=7cm]{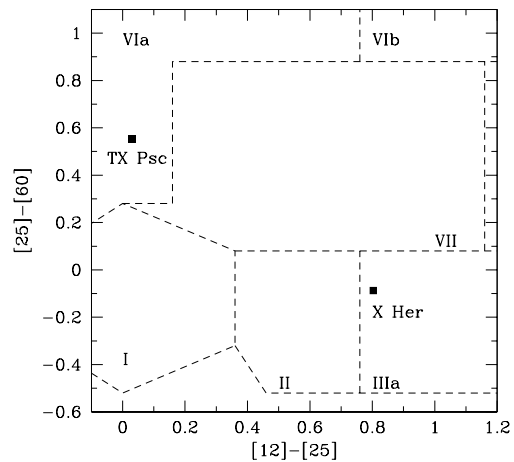}
     \caption{\label{Fig:IRAS}
The location of X~Her and TX~Psc in the IRAS colour-colour diagram, along with \citet{1988A&A...194..125V} regions.
}
\end{figure}


\section{Basic data of the targets}

\subsection{X~Her}

X~Her is an oxygen-rich M8III semi-regular variable (SRV) of period 95~d and with a long secondary period  of  746~d \citep{2002AJ....123.1002H}. From a radial-velocity monitoring of the CO lines at 1.6~$\mu$m, the same authors  discovered a period of 660~d. That all six but one among the SRV stars monitored by Hinkle et al. have similar eccentricities and longitudes of periastron casts doubts on the orbital origin of the observed velocity variations.  
The IRAS Low Resolution Spectrum of X~Her was classified as E (i.e., silicate emission) by \citet{1997ApJS..112..557K}, and as SE6t (i.e., showing a 13~$\mu$m feature, most probably due to Al-rich oxide dust,  and strong amorphous silicate bands) by 
\citet{1996ApJ...463..310S,2003ApJ...594..483S}. Its position in region IIIa of the IRAS colour-colour diagram shown in Fig.~\ref{Fig:IRAS} is consistent with the LRS class referring to a thick dust shell. From a model fit to the IRAS data, \citet{1993ApJS...86..517Y} found X~Her to be extended to a radius of $6.2\arcmin$ at 60~$\mu$m; however,  the published data (their Fig.~11) revealed
a much narrower profile, with a radius of the order of 2$\arcmin$, and some substructure, which could possibly be  attributed to the nearby pair of background galaxies  UGC~10156a and b (see Fig.~\ref{Fig:xher_red_blue} below).

\begin{figure}
\centering
   \includegraphics[width=7cm, height=7cm]{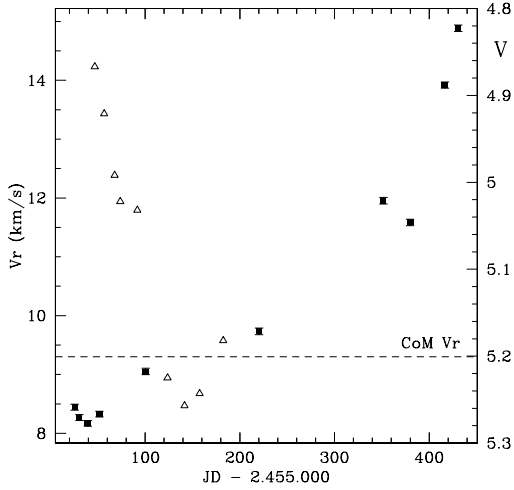}
     \caption{\label{Fig:Vr}
Radial velocities of TX~Psc obtained with the HERMES spectrograph  and an Arcturus cross-correlation mask (black squares and left scale), along with AAVSO $V$ magnitudes (open triangles and right scale).  The horizontal dashed line is the heliocentric centre-of-mass (CoM)  velocity from the CO $J=1-0$ radio line \citep[][and references therein]{1995AJ....110..805B}.
}
\end{figure}
\begin{table}
      \caption[]{Basic kinematical data for X~Her and TX~Psc. Parallaxes and proper motions are from \citet{2007A&A...474..653V}.
The column labelled 'LSR'  provides the kinematical data  corrected from the solar motion. }
      \label{pmdata}

      \begin{tabular}{lrrrr}          
\hline\hline      
& \multicolumn{2}{c}{X Her} & \multicolumn{2}{c}{TX Psc}\\
\hline\\
Spec. Type &    \multicolumn{2}{c}{M8 III} & \multicolumn{2}{c}{C5-7,2 (N)} \\
Var. type$^e$  &   \multicolumn{2}{c}{(95 d)} & \multicolumn{2}{c}{SR (224 d)}  \\
$l$    ($^\circ$)  &  \multicolumn{2}{c}{74.5} &   \multicolumn{2}{c}{93.3}\\
$b$    ($^\circ$)      &   \multicolumn{2}{c}{47.8} & \multicolumn{2}{c}{$-55.6$}\\
$\varpi\pm\sigma_\varpi$ (mas)  &  \multicolumn{2}{c}{ $7.30\pm0.40$} & \multicolumn{2}{c}{$3.63\pm0.39$}\\
\hline\\
& helioc. & LSR & helioc. & LSR\\
\hline\\
$\mu_\alpha \;\cos\delta$  (mas/yr)  & $-68.5\pm0.4$ &  $-57.3$ &   $-33.7\pm0.4$ & $-46.4$\\
$\mu_\delta$ (mas/yr)  & $64.6\pm0.4$  & 55.1&  $-24.5\pm0.3$ & $-19.0$\\
RV (km/s)  & -88.95$^{\;c}$  & $-73.6$&   13.0$^{\;a}$& 13.4\\
Syst. vel.  (km/s)  & -90$^{\;d}$ & &  9.3$^{\;b}$  \\
Space vel.  (km/s) 	& $108.0\pm2.0$ & 89.9 &  $55.2\pm4.4$ & 66.9\\ 
Pos. angle  ($^\circ$)   & 313.3 & 309.0 & 234.0 & 242.6\\
Inclination ($^\circ$) & & -55.4 & & 13.6 \\
\hline\\
\end{tabular}

References:
\\
a: Average HERMES velocity\\
b: CO $J = 1-0$ heliocentric velocity from \citet{1995AJ....110..805B} and references therein.\\
c: \citet{2005A&A...430..165F}\\
d: \citet{1996A&A...310..952K}\\
e: GCVS

\end{table}


\subsection{TX~Psc}

TX~Psc is one of the brightest and nearest optical carbon stars and has 
been extensively observed from the visual to mm-waves. The {\it General Catalogue of Variable Stars} (GCVS)  indicates a variability type Lb, but an average period of 224~d was found by \citet{1995IBVS.4159....1W,1997JAVSO..26....1W}, indicating that SRa/b may be more appropriate. The photometric variability in the $V$  band is of moderate amplitude (0.4~mag peak to peak in the $V$ band; see Fig.~\ref{Fig:Vr}). 
However, ten new 2009-2010 observations with the HERMES/MERCATOR spectrograph \citep{Raskin2010} spread over 450 days show large and regular variations in the radial velocity from 8 to 15 km~s$^{-1}$ (Fig.~\ref{Fig:Vr}); these radial velocities were derived by correlating the observed spectrum with a template mask containing 3100 lines covering the spectral range 476 -- 654~nm. Earlier velocity measurements by \citet{1995AJ....110..805B} confirm this amplitude range, but are too sparsely sampled to reveal any underlying regularity. Simultaneous AAVSO photoelectric measurements confirm the photometric period of about 220~d.
Given that the regular velocity variations occur on a period substantially longer than the SRa/b period of 224~d, these velocity variations cannot be interpreted as a manifestation of a pulsating atmosphere [see \citet{2001A&A...377..617L} for examples of a similar situation for radial velocities derived from 4~$\mu$m lines]. They could either be an indication of the possible binary nature of TX~Psc, or yet another example of the so-called {\it long-secondary periods} encountered in several long-period variables  \citep[including X~Her;][]{2002AJ....123.1002H,2009MNRAS.399.2063N} whose origin is still unclear. 
The radial-velocity monitoring is being pursued to clarify the origin of these variations. In the absence of any firm indication of the binary nature of TX~Psc, we do not consider that possibility any further in this paper, especially since it may just be of a similar (non-orbital) nature as that observed  for X~Her and discussed above.

Although the IRAS low resolution spectrum of TX~Psc has been classified as S (i.e., featureless, stellar-like) by \citet{1997ApJS..112..557K}, its 60 $\mu$m flux is clearly indicative of cool dust, TX~Psc falling in region~VIa of the IRAS colour-colour diagram (Fig.~\ref{Fig:IRAS}), which is  the locus of stars with detached dust shells. \citet{1993ApJS...86..517Y} indeed found TX~Psc to be extended with a radius of roughly 3$\arcmin$ at 60~$\mu$m. 
From a fit to the near-IR and IRAS spectral energy distribution, \citet{1986A&A...161..305E} have derived a dust temperature of 125~K, but a more elaborate study by \citet{1991ApJ...377..285W} suggests that the dust temperature decreases from 1155~K at 2.25 stellar radii to 410~K at 30 stellar radii.

\subsection{Kinematical data}

Proper motions, radial velocities, and parallaxes for X~Her and TX~Psc are listed in Table~\ref{pmdata}. For all calculations, we used  the revised Hipparcos data by \citet{2007A&A...474..653V}, and the systemic velocity derived from radio observations, which is consistent with the radial velocity derived from optical lines (Table~\ref{pmdata}). The space velocity is computed as follows. First 
the components $(U,V, W)$ of the spatial heliocentric velocity  in Galactic coordinates are computed  following \citet{1987AJ.....93..864J}, using the proper motion, the distance, and 
the radial velocity. The sources for these quantities used to compute the space velocity are given in Table~\ref{pmdata}. The solar motion  $(U,V, W)_0 = (11.10, 12.24, 7.25)$~km~s$^{-1}$ \citep{2010MNRAS.403.1829S} is then added, and the $\alpha$, $\delta$  components of the space velocity, thus corrected from the solar motion,  are then easily calculated by reversing the procedure.  In Table~\ref{pmdata}, the  column labelled 'LSR' provides all quantities corrected for the solar motion.

\begin{figure}
\centering
   \includegraphics[width=7cm]{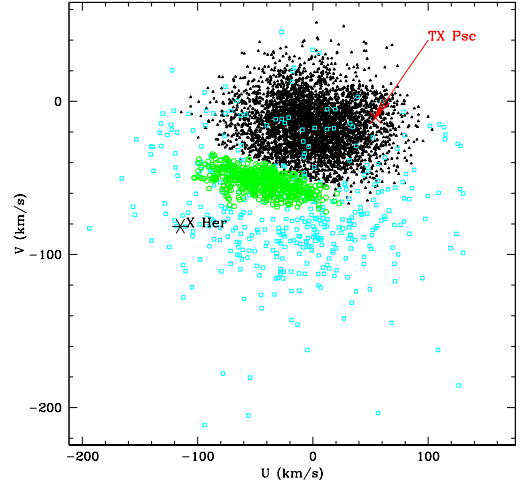}
     \caption{\label{Fig:UV}
The location of X~Her  and TX~Psc in the $UV$ diagram, with data for K and M giants taken from \citet{2005A&A...430..165F}. Black triangles denote stars belonging to the background velocity ellipsoid, (cyan) open squares to high-velocity stars, and (green) open circles to Hercules-stream stars.
}
\end{figure}

The location of  X~Her and TX~Psc and in the $UV$ diagram is shown in Fig.~\ref{Fig:UV}, which reveals that X~Her belongs to the group flagged as 'high-velocity stars' by \citet{2005A&A...430..165F}. TX~Psc, in contrast, happens to fall in the background ellipsoid.

\section{Herschel data analysis}
\label{Sect:data_analysis}

\subsection{Observations and data reduction}
\label{observation}
\begin{figure}[t]
\centering
   \includegraphics[width=7cm]{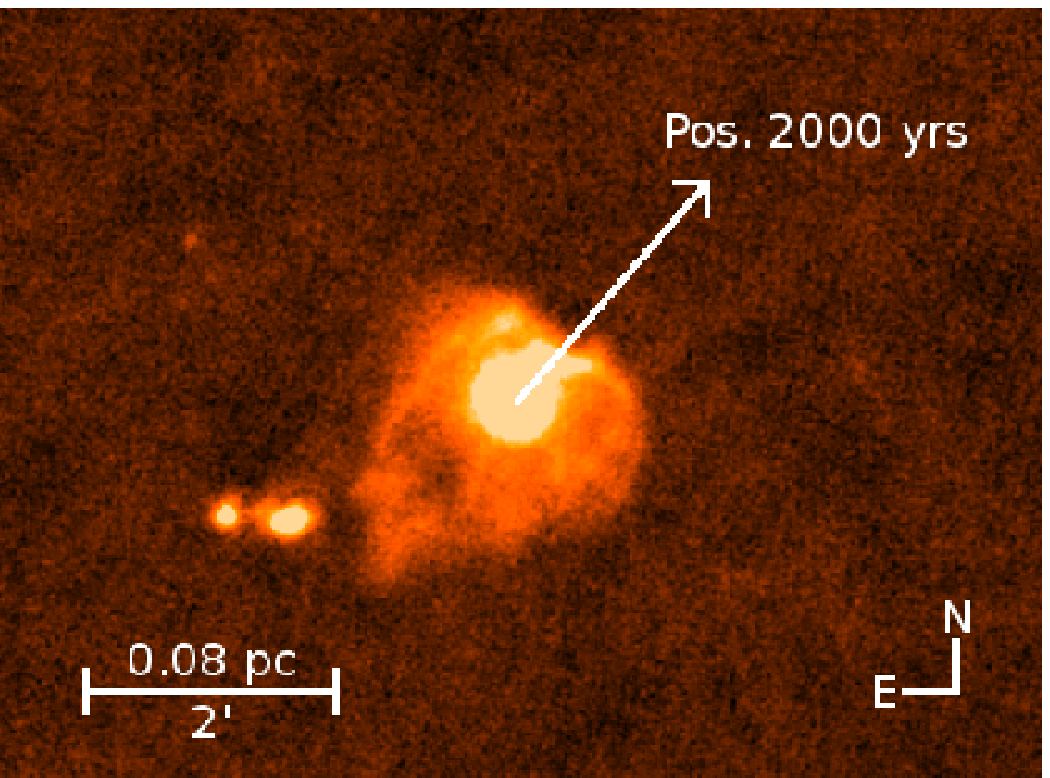}
   \\
   \includegraphics[width=7cm]{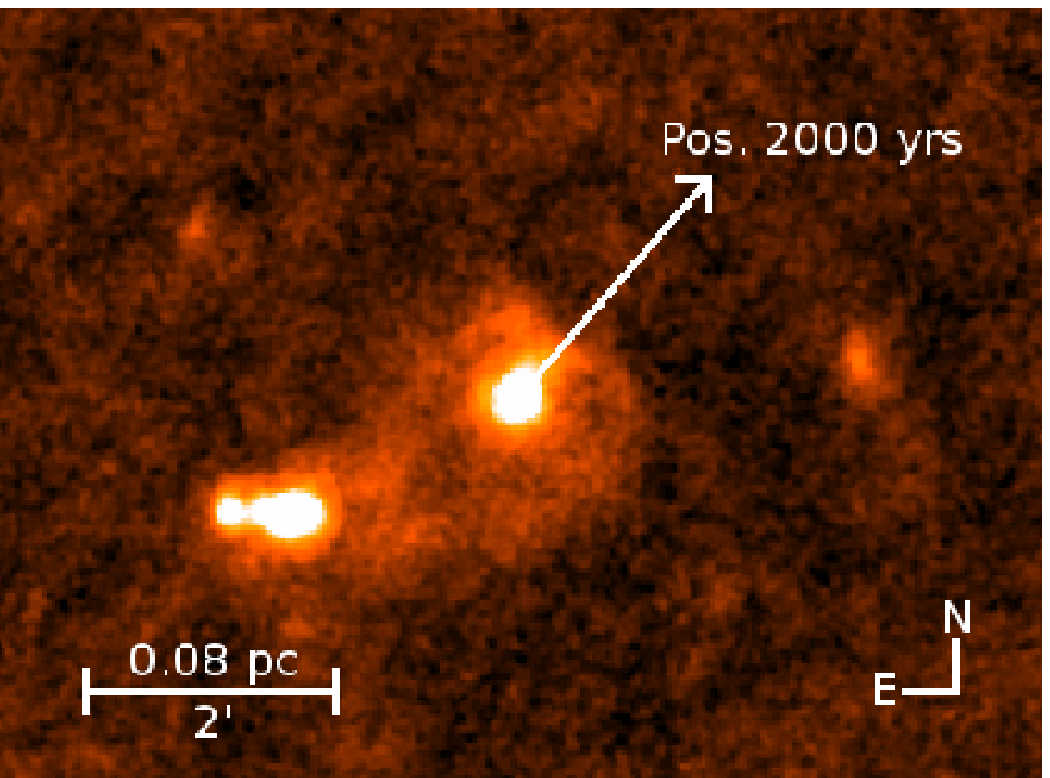}
     \caption{\label{Fig:xher_red_blue}
PACS images of X~Her. The upper image is at 70~$\mu$m, the lower at 160~$\mu$m. The arrows indicate the direction of the proper motion and  the position 2000 yrs in the future. The two bright dots on the lower left side are a pair of galaxies (UGC 10156a and b).   The faint source west of X~Her is the galaxy 
MCG+08-29-036.}
\end{figure}

\begin{figure}[t]
\centering
   \includegraphics[width=7cm]{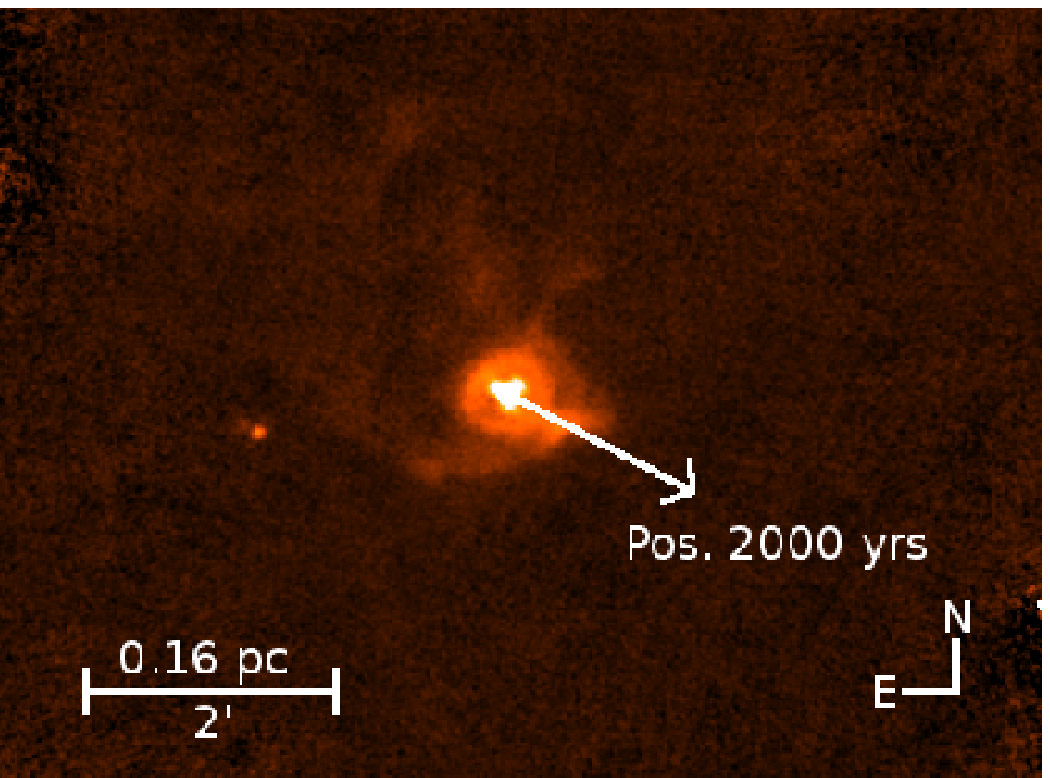}
   \\
   \includegraphics[width=7cm]{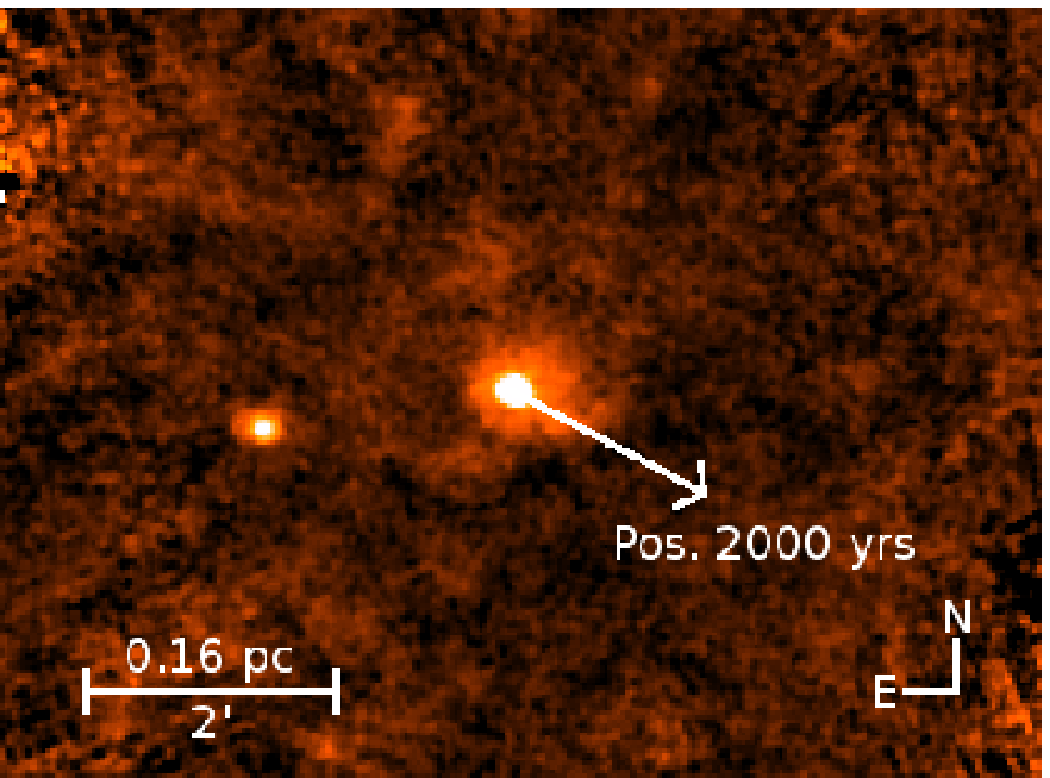}
     \caption{\label{Fig:txpsc_red_blue}
Same as Fig. \ref{Fig:xher_red_blue} for TX~Psc. The bright source east of the star is an uncatalogued galaxy.
}
\end{figure}

The images of X~Her (Fig.~\ref{Fig:xher_red_blue}) and TX~Psc (Fig.~\ref{Fig:txpsc_red_blue}) that we present here were obtained by the Photodetector Array Camera and Spectrometer (PACS) of Herschel on 2009 December 20 and 21, respectively, and are part of the Mass-loss of Evolved StarS (MESS) guaranteed time key programme. Details of the MESS programme can be found in \citet{Groenewegen_2011}, which describe as well the basic data processing and reduction with the Herschel Interactive Processing Environment (HIPE) that was used for producing Figs.~\ref{Fig:xher_red_blue} and \ref{Fig:txpsc_red_blue}. 
All images were oversampled by a factor of 3.2, which results in a sampling of 1~pixel per arcsec at 70~$\mu$m and 1~pixel per 2~arcsec at 160~$\mu$m, to be compared with the  FWHM of $5\farcs7$  and $11\farcs4$ at those wavelengths, respectively. Fig.~\ref{Fig:txpsc_xher_deconv} presents deconvolved images of X~Her and TX~Psc. More details of the deconvolution methods used to produce these images can be found in \citet{ottensamer_galagb}.

\subsection{Structure and dynamical  age of the circumstellar nebula}
\label{Sect:blob}

\begin{figure*}
\centering
   \includegraphics[width=9cm]{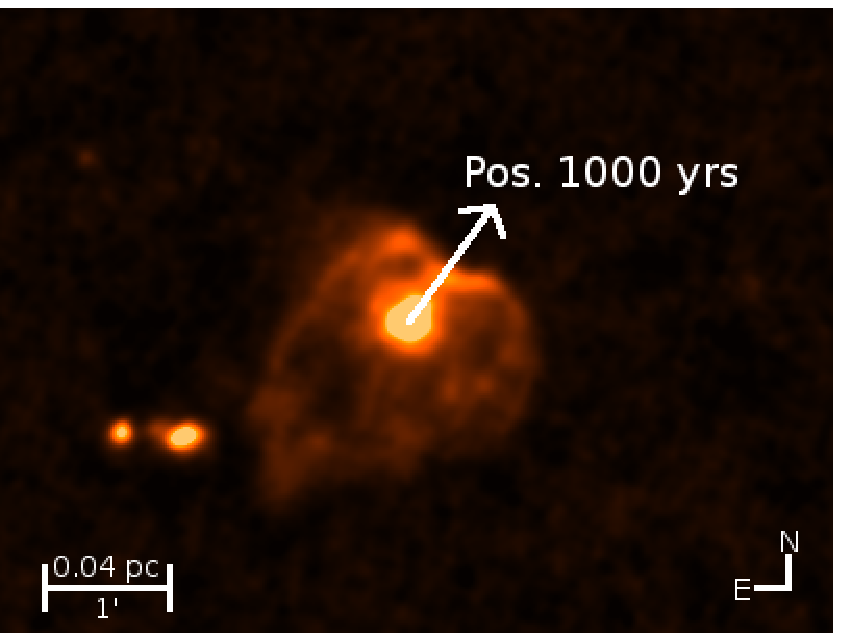}
   \includegraphics[width=9cm]{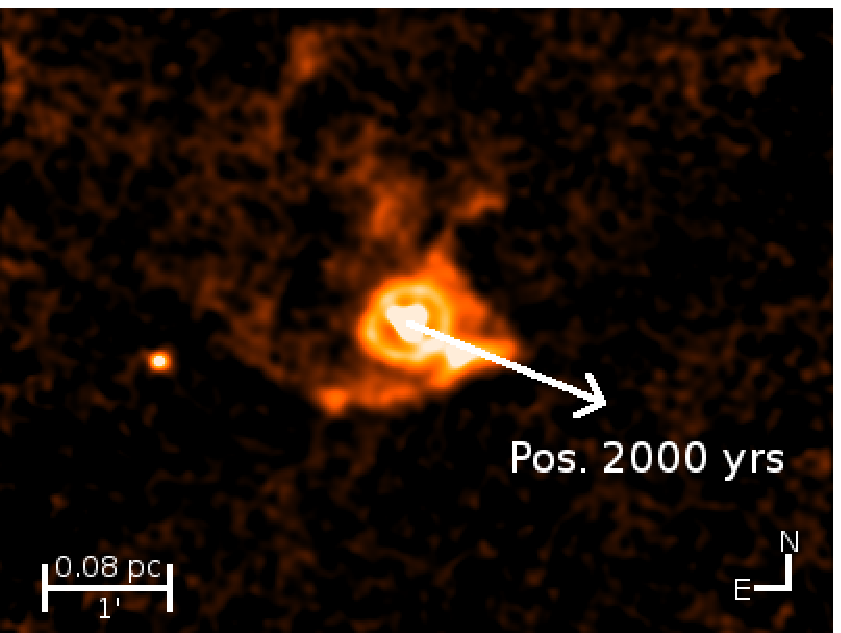}

     \caption{\label{Fig:txpsc_xher_deconv}
Deconvolved PACS images of  X~Her (left) and TX~Psc (right), both in the blue channel at 70~$\mu$m. The arrows indicate the direction of the proper motion and the position 1000 or 2000~yrs in the future. North is up and east is to the left.
}
\end{figure*}

The images of X~Her (Figs.~\ref{Fig:xher_red_blue} and \ref{Fig:txpsc_xher_deconv}) and TX~Psc (Figs.~\ref{Fig:txpsc_red_blue} and \ref{Fig:txpsc_xher_deconv}) reveal a wealth of features that we now discuss.

For both  X~Her and TX~Psc,  there is a noteworthy clump  in
the direction of the space motion (Fig.~\ref{Fig:txpsc_xher_deconv} and bottom right panel of Fig.~\ref{Fig:txpsc_scan}). 
For TX~Psc, there 
is also a ring around the star, which is clearly visible in the across scan cut (upper panel of Fig.~\ref{Fig:txpsc_scan}). 
This ring is not an artefact of the deconvolution process, since it is also visible in the non-deconvolved image (Fig.~\ref{Fig:txpsc_red_blue}). Moreover,  a similar circular extended (filled) structure is also visible in the 70~$\mu$m image  of X~Her (Fig.~\ref{Fig:txpsc_xher_deconv}). TX~Psc is close to the centre of this ring (the upstream radius is the same as the downstream radius; see Fig.~\ref{Fig:txpsc_scan}). The circular symmetry of this feature is a clear indication that the mass loss from which it originates was  isotropic, at variance with the situation currently prevailing closer to the star, where a bipolar outflow is observed (see Sect.~\ref{Sect:bipolar}). Moreover, the prominent nature of this ring on the PACS images may indicate that it actually corresponds to the termination shock, where the reverse shock from the wind-ISM interface is facing the freely expanding stellar wind, although the available data offer so far no additional support of this hypothesis.

The dynamical ages of these features may be estimated as follows. The inner ring is at a distance of $17\arcsec$ from TX~Psc downstream, which corresponds to a kinematic age of about $2100$~yrs (Fig.~\ref{Fig:txpsc_scan})
for an assumed wind velocity $v_w = 10.5$~ km~s$^{-1}$ \citep{1993ApJS...87..267O}, neglecting any possible inclination on the plane of the sky. 
The blob in the direction of the space motion is at a distance of $29\arcsec$, corresponding to a  kinematic age of $3610\pm410$~yrs. 
The  wind-ISM interface limits the possibilityof looking further back into the  mass-loss history.

\begin{figure}
\centering
   \includegraphics[angle=-90,width=7cm]{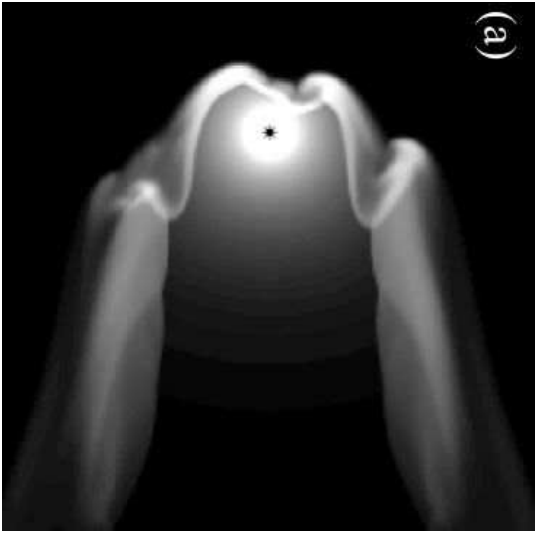}
     \caption{\label{Fig:Wareing}
The gas density in case E simulation of \citet{2007MNRAS.382.1233W} (corresponding to $v_{*,{\rm ISM}}=125$~km~s$^{-1}$,  $n_H=2$~cm$^{-3}$, and a mass-loss rate of $5\;10^{-6}$~M$_\odot$~yr$^{-1}$), which reveal the formation of vortices in the wake of the AGB star.
This stage corresponds to the end of the AGB phase, and resembles the X~Her nebula (compare with Fig.~\ref{Fig:txpsc_xher_deconv}, remembering that the viewing angle is different, since the space motion of X~Her is inclined by 55$^\circ$ with respect to the plane of the sky). 
Reprinted from MNRAS, Volume 382, Wareing et al., page 382, 2007, with permission from John Wiley \& Sons.
}
\end{figure}
\begin{figure}
\centering
   \includegraphics[width=5cm]{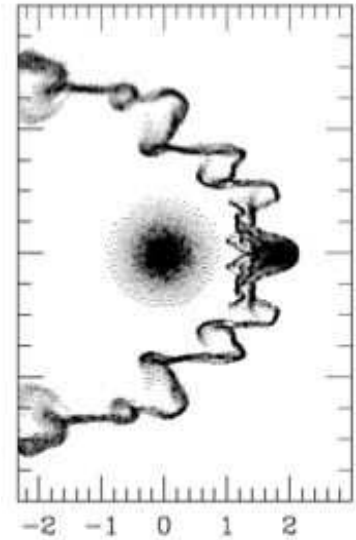}
     \caption{\label{Fig:Blondin}
The gas density in the Mach 10 simulation of \citet{1998NewA....3..571B}. Note the strong blob upstream and the highly unstable and inhomogeneous bow shock. This case resembles the TX~Psc nebula (compare with Fig.~\ref{Fig:txpsc_xher_deconv}, remembering that the viewing angle is slightly different, since the space motion of TX~Psc is inclined by 14$^\circ$ with respect to the plane of the sky). Reprinted from New Astronomy
Volume 3, John M. Blondin and Joel F. Koerwer, pages 571-582, 1998,  with permission from Elsevier.
}
\end{figure}
\begin{figure}[t]
\centering
\begin{minipage}[t]{9cm}
   \includegraphics[width=4.4cm]{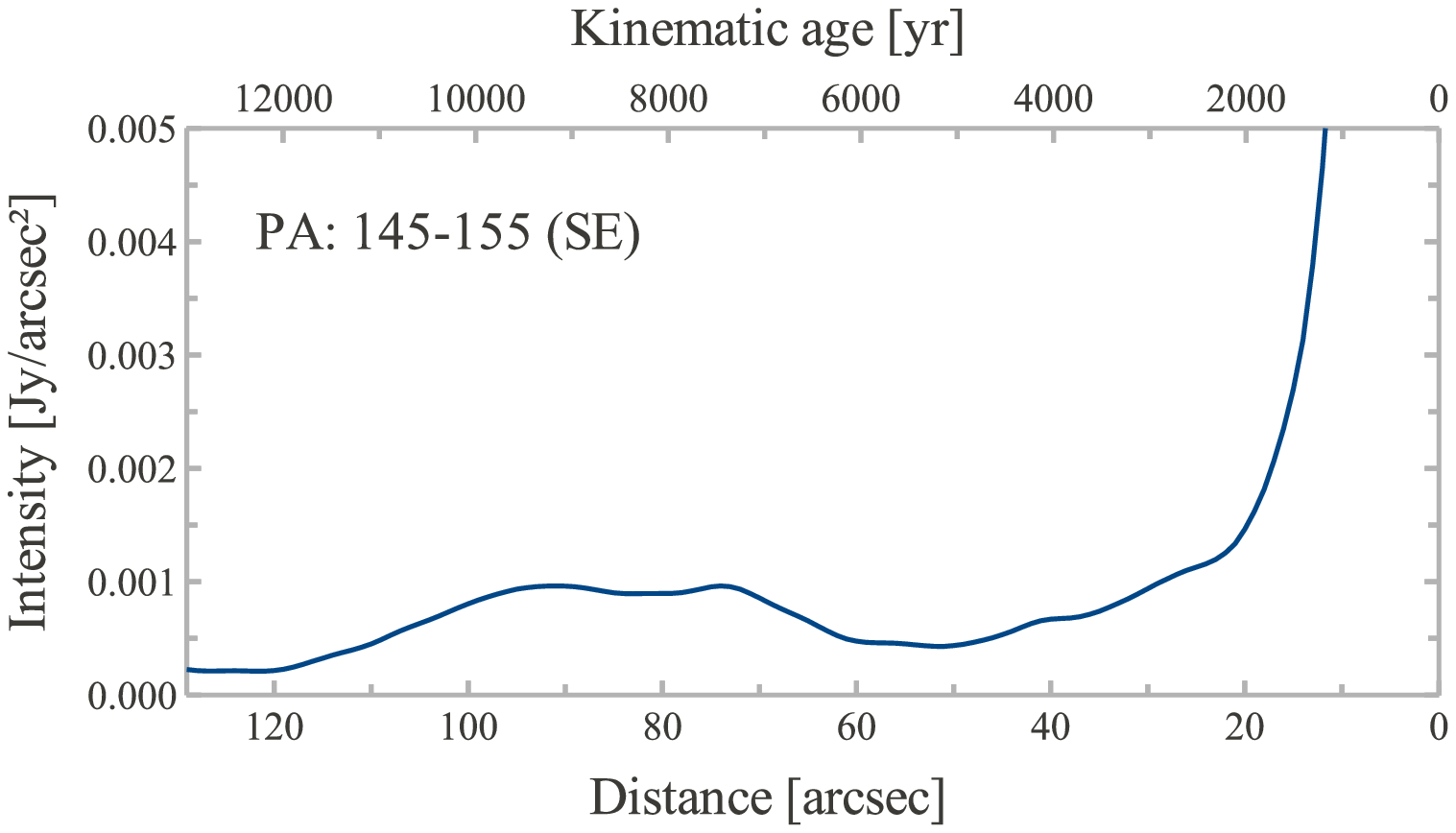}
   \includegraphics[width=4.4cm]{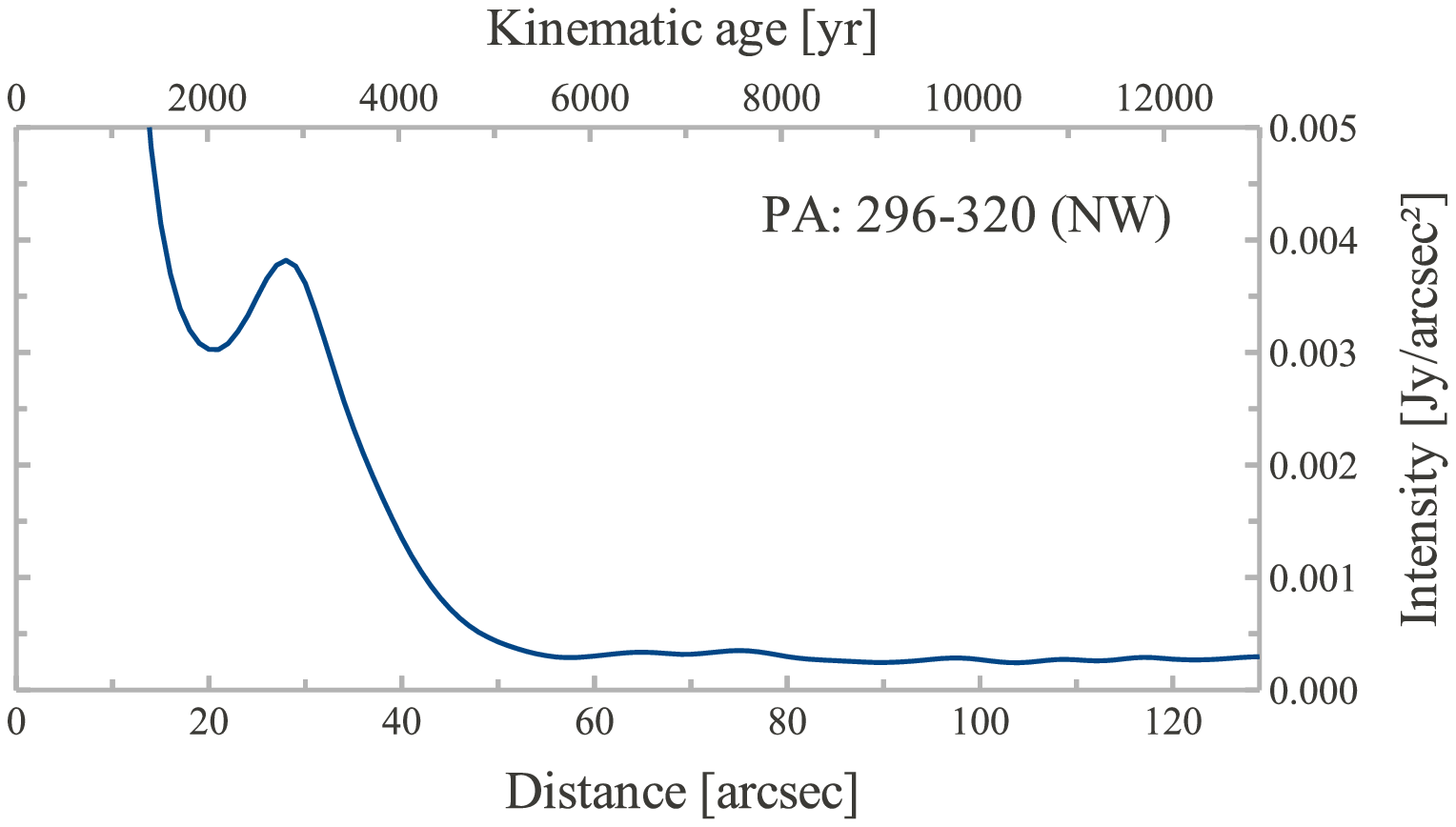}
\end{minipage}
\\
\begin{minipage}[t]{9cm}
   \includegraphics[width=4.4cm]{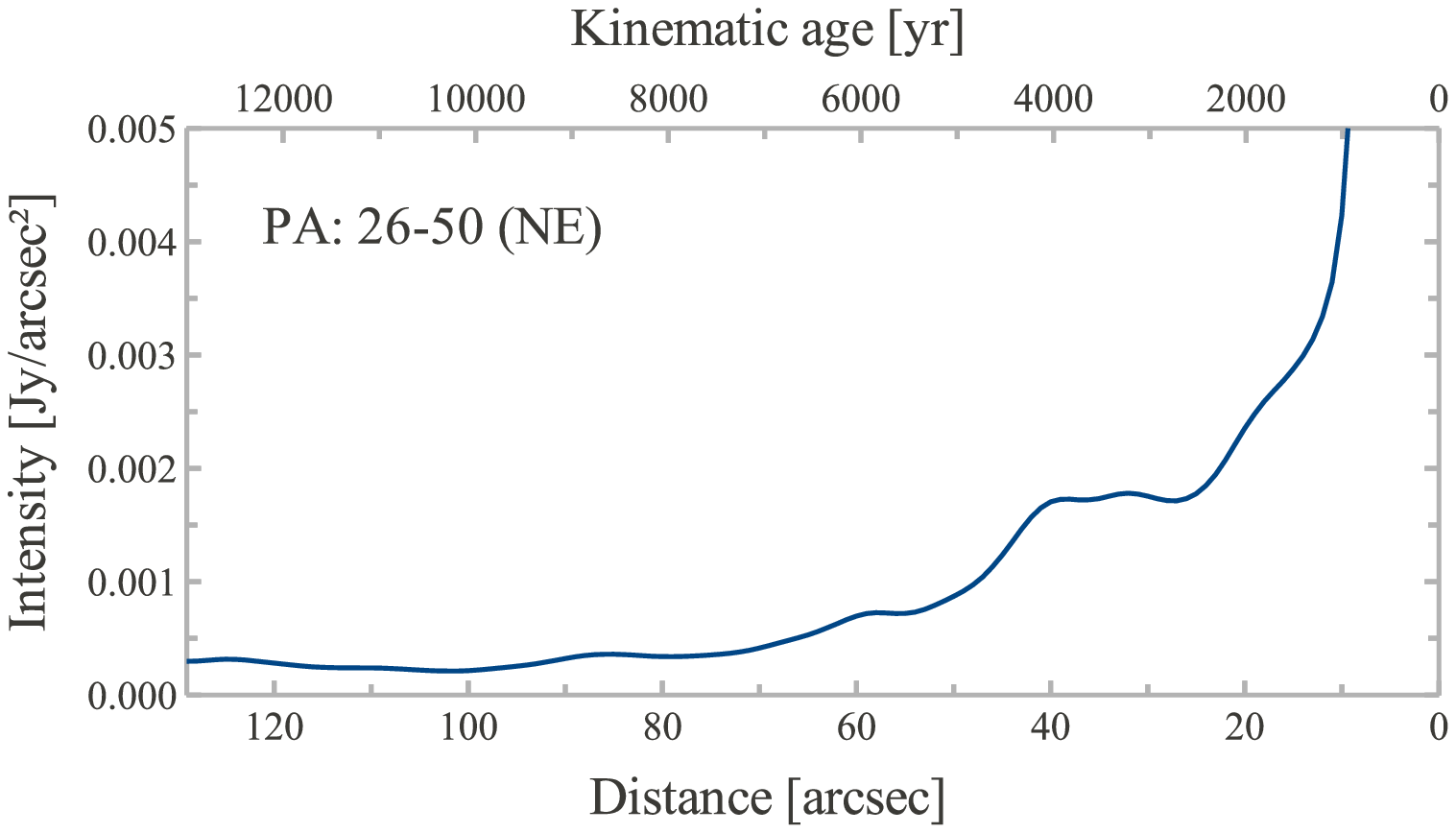}
   \includegraphics[width=4.4cm]{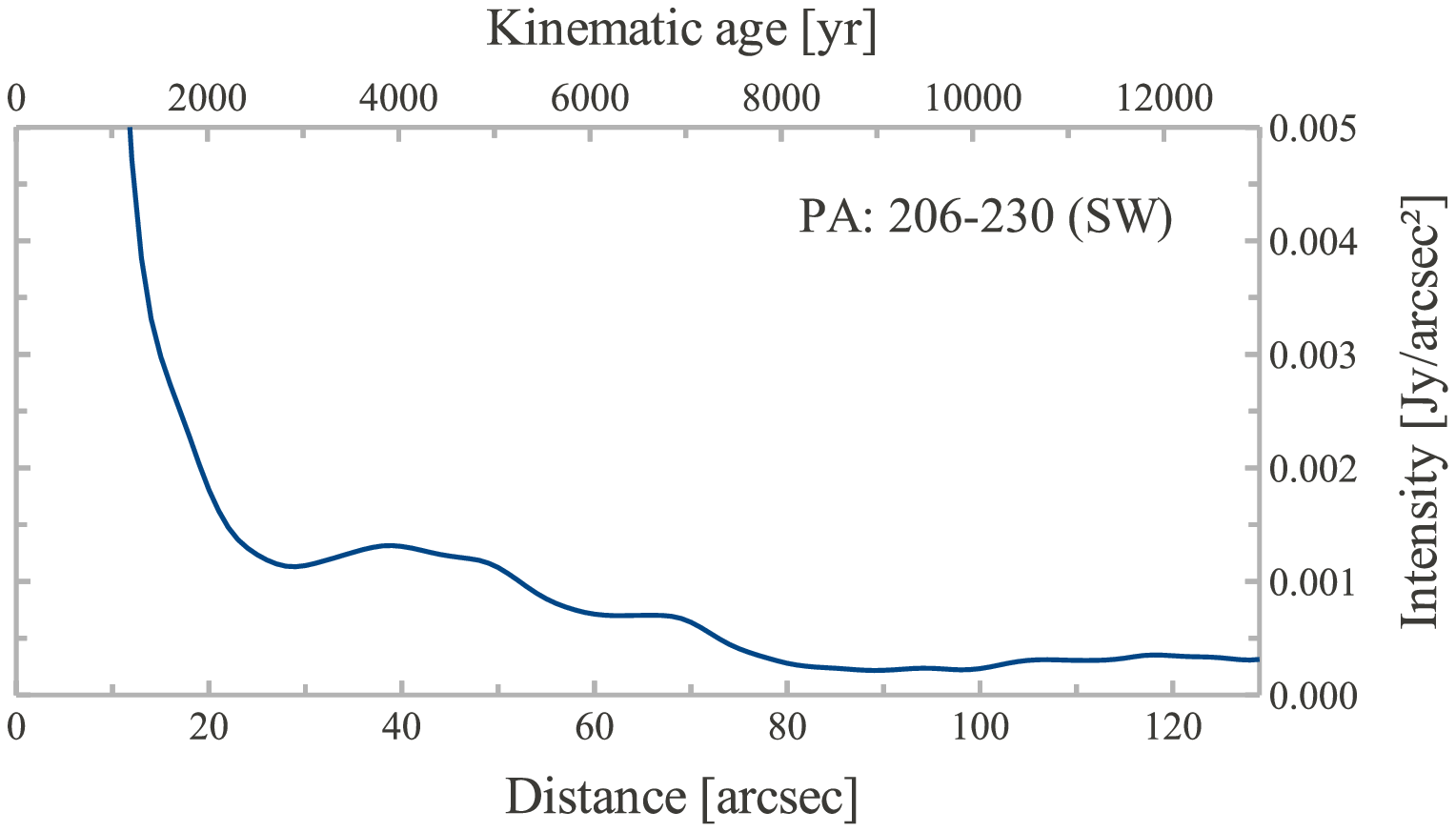}
\end{minipage}
     \caption{\label{Fig:xher_scan}
Intensity cuts for X~Her from the PACS $70$~$\mu$m deconvolved image (right panel of Fig.~\ref{Fig:txpsc_xher_deconv}) along two perpendicular directions through the star, one corresponding to the space motion direction (lower panels). The dynamical ages on the upper scale correspond to a wind expansion velocity of 6.5~km~s$^{-1}$. 
}
\end{figure}
\begin{figure}[t]
\centering
\begin{minipage}[t]{9cm}
   \includegraphics[width=4.4cm]{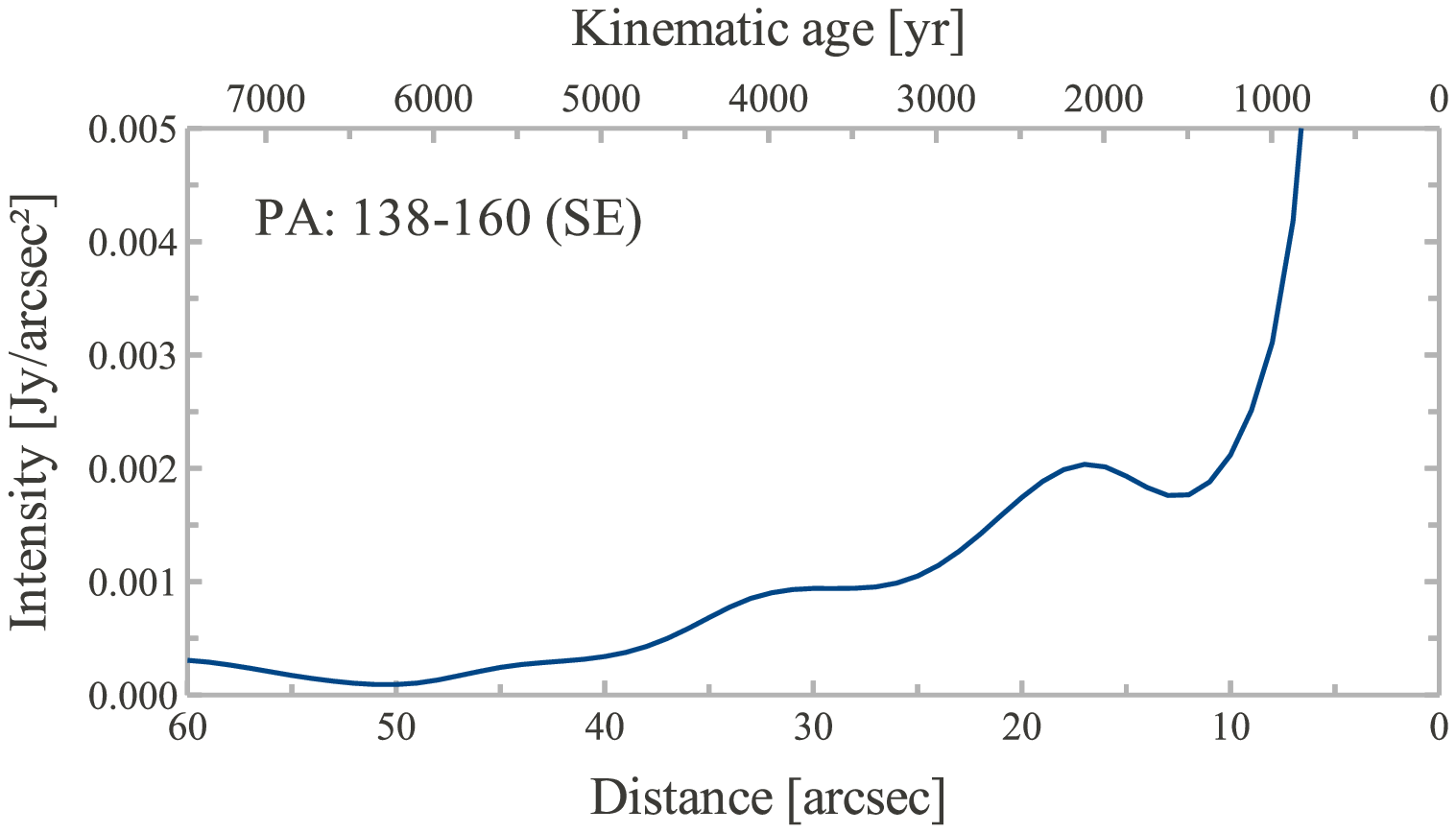}
   \includegraphics[width=4.4cm]{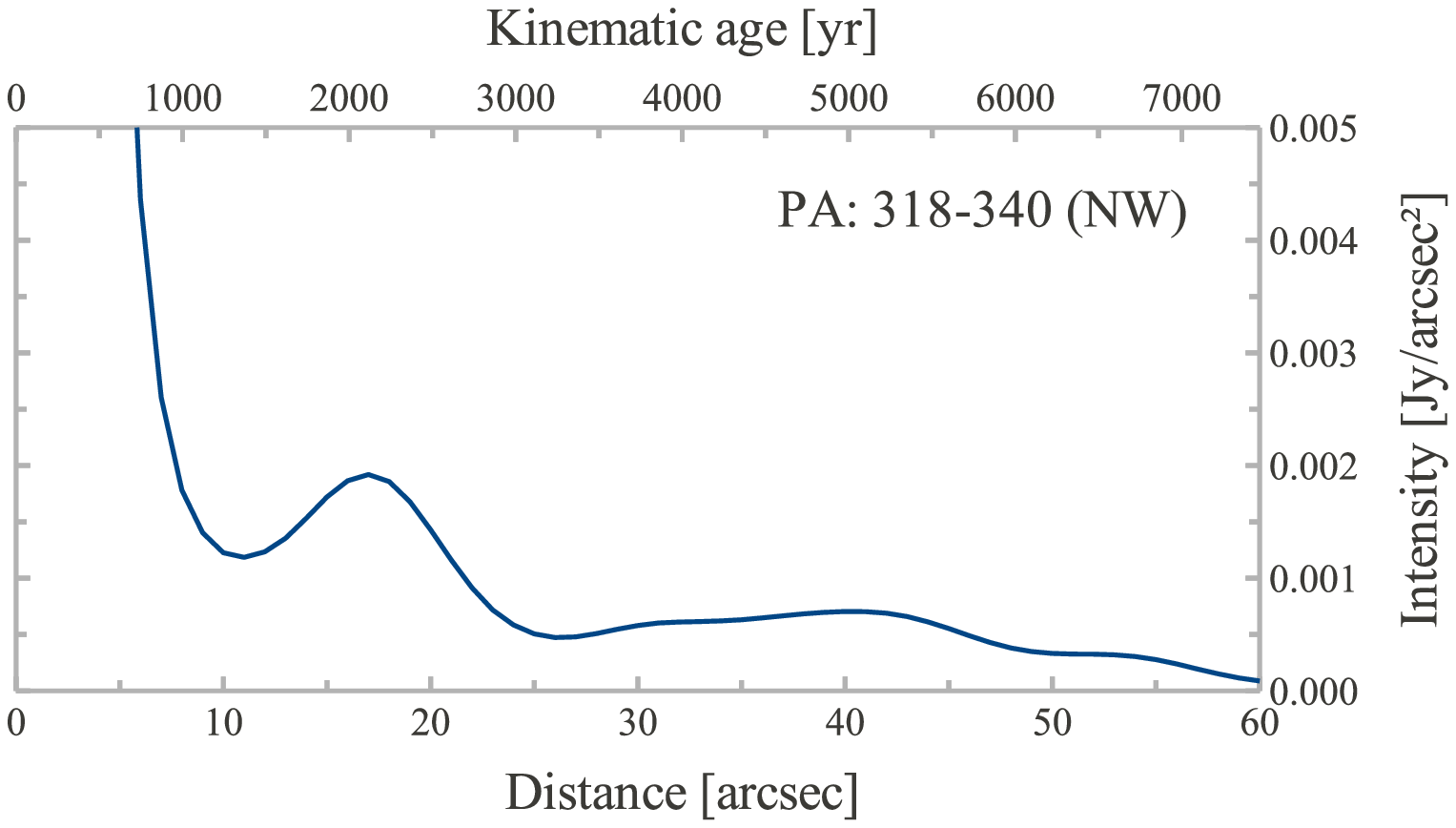}
\end{minipage}
\\
\begin{minipage}[t]{9cm}
   \includegraphics[width=4.4cm]{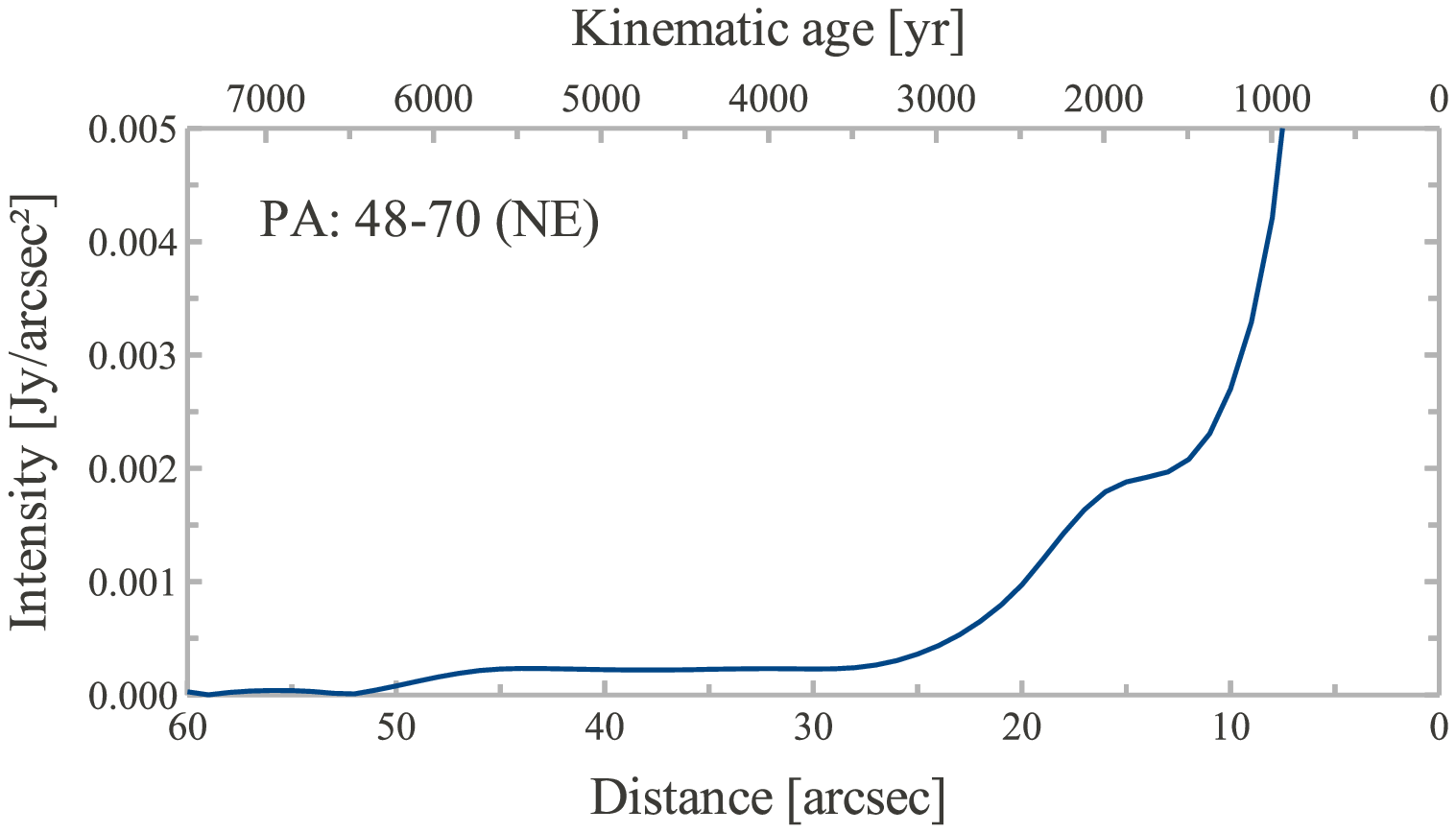}
   \includegraphics[width=4.4cm]{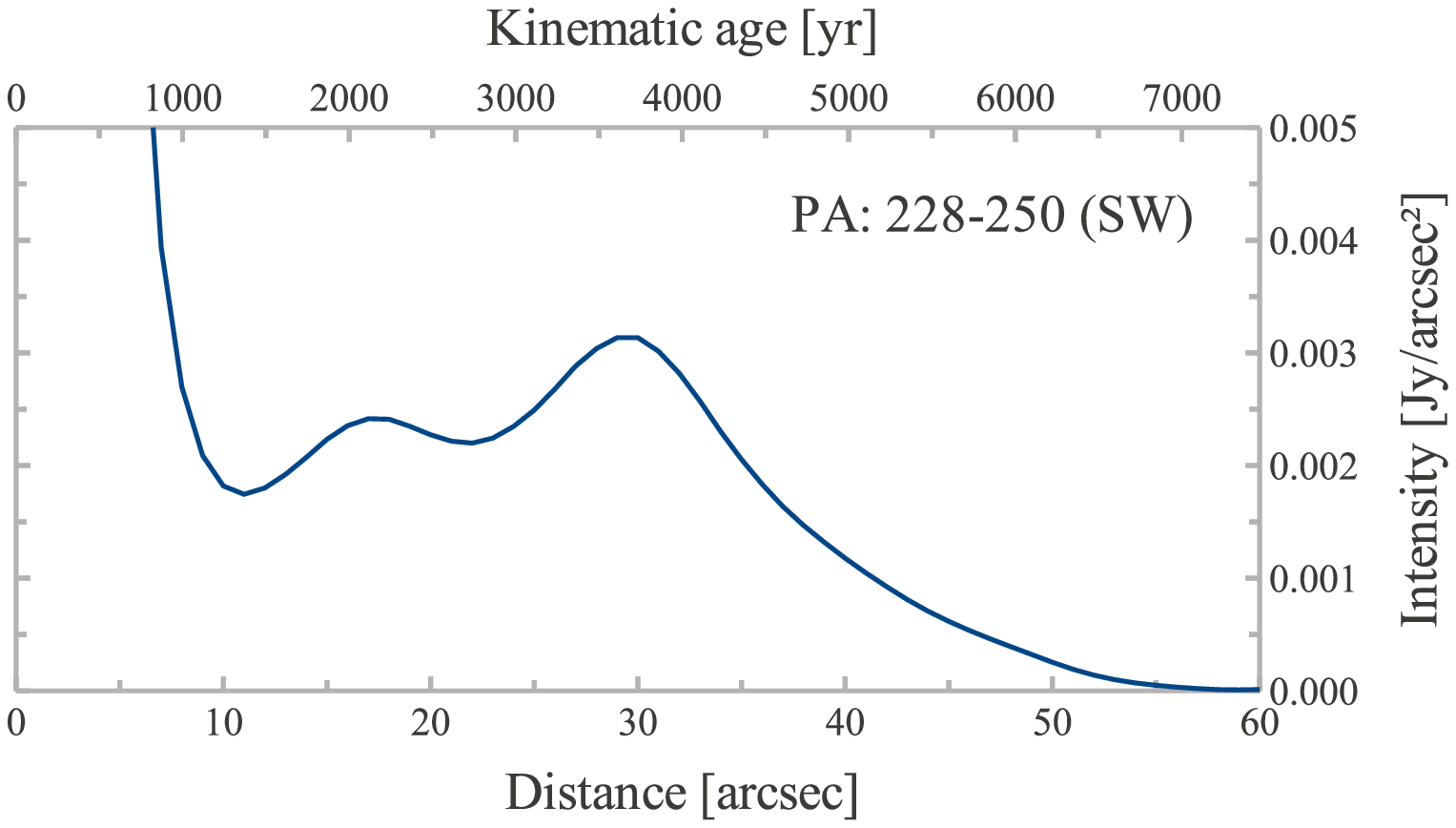}
\end{minipage}
     \caption{\label{Fig:txpsc_scan}
Same as Fig.~\ref{Fig:txpsc_scan} for TX~Psc and  a wind expansion velocity of 10.5~km~s$^{-1}$.
}
\end{figure}

The same procedure and assumptions were applied to the heart-shaped X~Her. With a wind velocity $v_w = 6.5$~km~s$^{-1}$ \citep{2003A&A...411..123G}, the bow shock at a distance of 28$\arcsec$ from the star has a minimum kinematic age of $2800\pm180$~yrs and the end of the tail at 112$\arcsec$ downstream has a (minimum) age of $11200\pm680$~yrs (Fig.~\ref{Fig:xher_scan}).

Another structural feature in TX~Psc deserves discussion. 
A compact feature (representing 2\% of the flux in the $K$ band and located upstream, at a position angle of 241$^\circ$) was  detected by the COME-ON+ observations of \citet{1998A&A...338..132C}, but at a distance of  only $0\farcs35$ from  the primary source, thus well inside the ring described above.
Although, at first sight, one would be tempted  to relate this blob observed close to the star with 
the one observed in the PACS 70~$\mu$m image discussed above, they cannot be caused by the same physical phenomenon (such as  the hydrodynamical instabilities discussed later in the present section), because the COME-ON+ secondary source is located {\it within}\footnote{It is in principle possible that this secondary source is actually located {\it outside} the circular ring, but seen inside only in projection. This  would mean that the secondary source is then located almost along the line of sight, at a distance larger than about 4700~AU ($=17\arcsec/0\farcs0036$) from the star. In the absence of any other argument supporting this view, this possibility will not be  considered further in this paper.} the circular ring observed around TX~Psc at a distance of about 17$\arcsec$. As discussed earlier, 
structures within this ring cannot be caused by the interaction with the ISM.
Nevertheless, the inner clump is observed at the same position angle as that corresponding to the space motion. Another puzzle, possibly related to the former, is the correlation existing between the space motion and the orientation of the bipolar outflows observed around
both X~Her (where it is almost exactly perpendicular to the space motion) and TX~Psc (where it seems to be parallel with the space motion). 
The properties of the bipolar outflow are discussed further  in Sect.~\ref{Sect:bipolar}. Finally, it is rather surprising that the morphology of the far-IR nebula does not bear any signature of the inner bipolar  outflow, especially in the case of X~Her where it is perpendicular to the space motion.

\citet{1998A&A...338..273C}  illustrated the wide variety of bow-shock structures that may arise from the interaction of the stellar wind with the ISM. They stressed the importance of the wind parameters (density and velocity), the star's velocity, and the ISM density. These parameters in turn fix the cooling efficiency of the gas, which appears to be an essential parameter in fixing  
the structure of the wind -- ISM interface \citep{1991AJ....102.1381S,1993ASPC...35..315V,1998A&A...338..273C}: if the cooling is efficient\footnote{See \citet{1991AJ....102.1381S} for a basic description of the relation between the wind/ISM properties and the cooling efficiency. Dense gas generally gives rise to efficient cooling.}, the gas can be compressed to high densities, and the shocked layer remains very thin\footnote{Wilkin fitting -- geometrical fitting of the bow shock shape based on the analytical formulation by \citet{1996ApJ...459L..31W}, which will be attempted in Sect.~\ref{wilkin} -- applies to the thin-shell approximation.}. This situation is described well by the isothermal approximation. On the contrary, if cooling is inefficient, the high temperature of the gas limits the compression factor, which for the limiting case of no cooling, would be equal to 4 \citep[abiabatic shock;][]{1998A&A...338..273C}.  In this situation, the shocked layer is thick. Different instabilities are encountered in thin and thick shocks \citep[see][for a review]{1998A&A...338..273C}. 

In thin shells, \citet{1996ApJ...461..372D,1996ApJ...461..927D} studied the so-called \textit{transverse acceleration instability}, which occurs as a consequence of the acceleration in the flow normal to the shock caused by its curved surface. Another instability appearing under the same conditions is the \textit{non-linear thin-shell instability}, which arises from the shear in shock-bounded slabs produced by deviations from equilibrium \citep{1994ApJ...428..186V,
1996NewA....1..235B}. If there is a shear along the surface of the bow shock, Kelvin-Helmholtz instabilities will develop as well.  
Numerical simulations relative to bow shocks with thin shells were performed by \citet{1991AJ....102.1381S}, \citet{1996ApJ...461..372D}, \citet{1998A&A...338..273C}, and \citet{1998NewA....3..571B}.

If the shocked material cools rather inefficiently, a thick shell will form, which will be subject to other types of instabilities, mostly the Rayleigh-Taylor (R-T) instability, because the densities of the gas entering from either side of the shock will not match in the absence of cooling. When two fluids of different densities are penetrating each other, R-T instabilities\footnote{A related instability is the Richtmyer-Meshkov instability \citep[e.g.,][]{Needham2010}, which develops when   a shock wave hits the interface of two fluids with different densities, such as  when the new shell of an expanding wind bubble hits the already established bow shock. This may happen as AGB stars lose mass in a succession of discrete events, as suggested by the successive shells observed around e.g. the carbon Mira IRC +10216 \citep{2000A&A...359..707M}.} manifest themselves in the form of mushroom- and finger-shaped protuberances of the lighter fluid into the heavier one. \citet{1991AJ....102.1381S} demonstrated that  R-T instabilities occur  when a stellar wind or planetary nebula shell is colliding with a low-density ISM with a  high relative velocity ($v_{*,{\rm ISM}} > 100$~km~s$^{-1}$), leading to a fragmentation of the shell. 
A bump in the upstream direction, attributed to a R-T instability developing at an early stage of the wind -- ISM interaction,  is seen for instance in the planetary nebula NGC 40 \citep{2002A&A...391..689M}.
At a later stage, as the shell fragments, Kelvin-Helmholtz (K-H) instabilities appear as a sequence of vortices peeling off the bow shock and moving downstream \citep{2007MNRAS.382.1233W,2007ApJ...660L.129W}.  

The nebulae around X~Her and TX~Psc  show very unusual features, such as the upstream blob and clumps visible along what seems to be a bow shock. 
Investigations of the shape of the bow shock  have been made using hydrodynamical simulations by  \citet{1998NewA....3..571B}, \citet{1998A&A...338..273C}, \citet{2003ApJ...585L..49V}, and \citet{2007MNRAS.382.1233W}.
The Case~E simulation of the last set of authors (corresponding to $v_{*,{\rm ISM}}=125$~km~s$^{-1}$,  $n_H=2$~cm$^{-3}$, and a mass-loss rate of $5\times10^{-6}$~M$_\odot$~yr$^{-1}$; see Fig.~\ref{Fig:Wareing}), at stage a (corresponding to the end of the AGB phase), 
looks very similar\footnote{We stress, however, that Wareing et al. numerical models do not incorporate dust grains, and only handle gas warmer than $10^4$~K (because of the lack of an adequate cooling function for gas with lower temperatures). In such environments, dust grains would not be able to survive and therefore far-IR dust thermal emission is not expected. Although Fig.~\ref{Fig:Wareing} shows structures resembling what is seen in PACS images, this model traces the hot gas density, which is not expected  to emit the observed far-IR emission. Despite these shortcomings,  the striking similarity beween the PACS and model images could be accounted for if the far-IR radiation comes from atomic emission lines in shocked gas. Alternatively, one may speculate that future models including the proper gas cooling function  will find similar structures involving gas at lower temperatures, with dust naturally forming  in regions of high gas densities  and moderate temperatures.} to the X~Her nebula (compare Figs.~\ref{Fig:txpsc_xher_deconv} and \ref{Fig:Wareing}), since it combines an upstream shock surface bent \textit{towards} the star, and K-H vortices in the downstream wake. Interestingly, the 90~km~s$^{-1}$ LSR space velocity 
of X~Her (Table~\ref{pmdata}) is similar to the velocity used in the Wareing et al. simulation, but the \textit{current} mass-loss rate is a factor of at least 30 lower than the Wareing value (see Sect.~\ref{Sect:XHer}).  On the other hand, the TX~Psc nebula bears similarities (especially the strong upstream blob, and the inhomogeneous bow) with the large Mach number (10) and isothermal simulations of Blondin \& Koerwer (1998; compare Figs.~\ref{Fig:txpsc_xher_deconv} and \ref{Fig:Blondin})\nocite{1998NewA....3..571B}. A large Mach number is indeed expected in the present situation, since the X~Her space velocity is 67~km~s$^{-1}$ and the sound speed is of the order of 1~km~s$^{-1}$ (for a temperature of the order of 80~K, as inferred from the analysis of Sect.~\ref{dusttemp}).
The nature of the ring observed in the TX~Psc image is unclear. Similar rings are clearly visible in the Wareing et al. simulations when the mass-losing star has reached the post-AGB phase (thus posterior to the 
stage displayed  in Fig.~\ref{Fig:Wareing}). The ring thus corresponds  to the planetary nebula shell. This cannot be the case for TX~Psc, although the ring possibly traces a discrete episode of strong mass loss. Alternatively, it could be the termination shock.

\subsection{Interpreting the emission of the upstream blob}
\label{dusttemp}

\begin{figure}[t]
\centering
   \includegraphics[width=7.5cm]{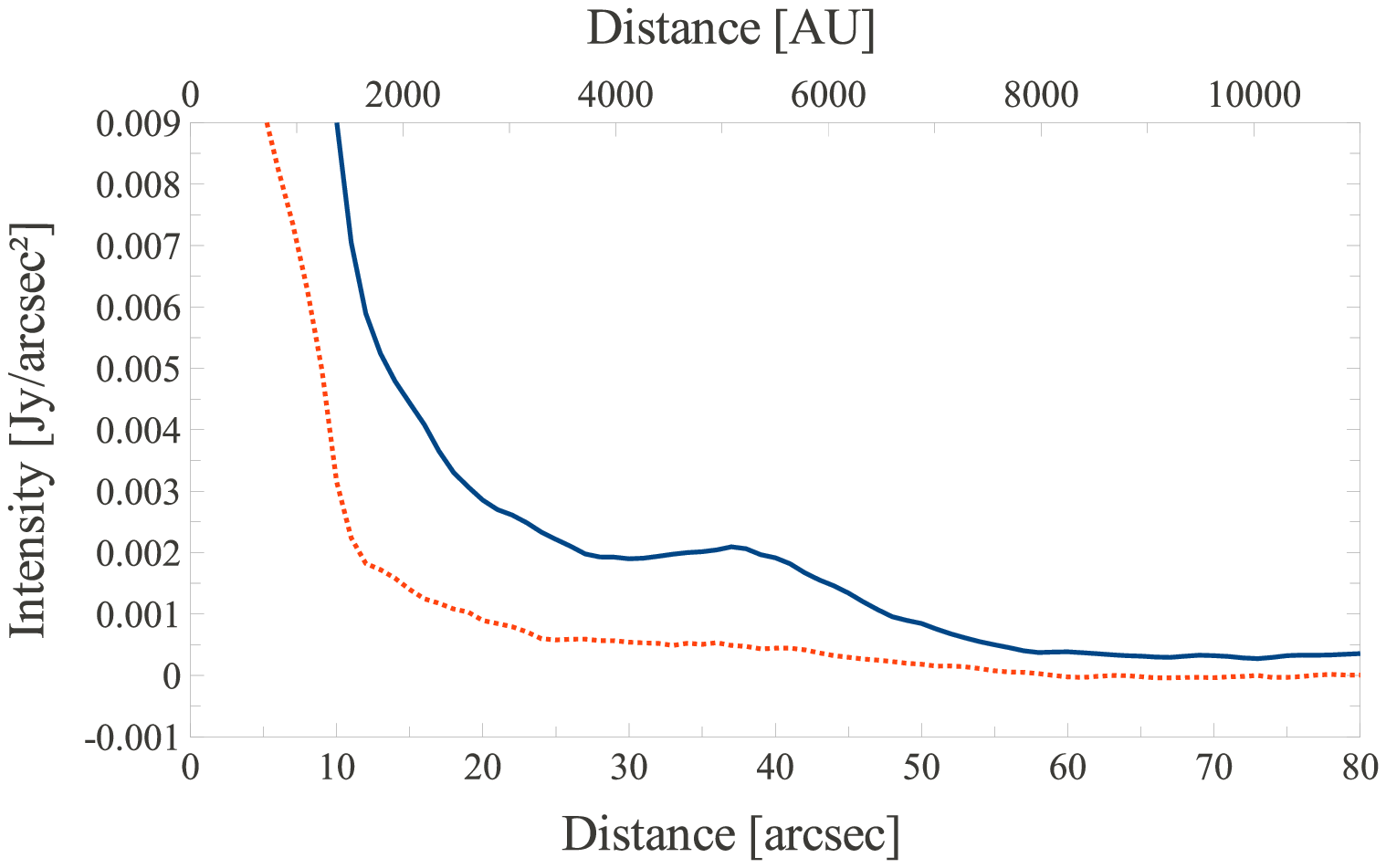}
\\
   \includegraphics[width=7.5cm]{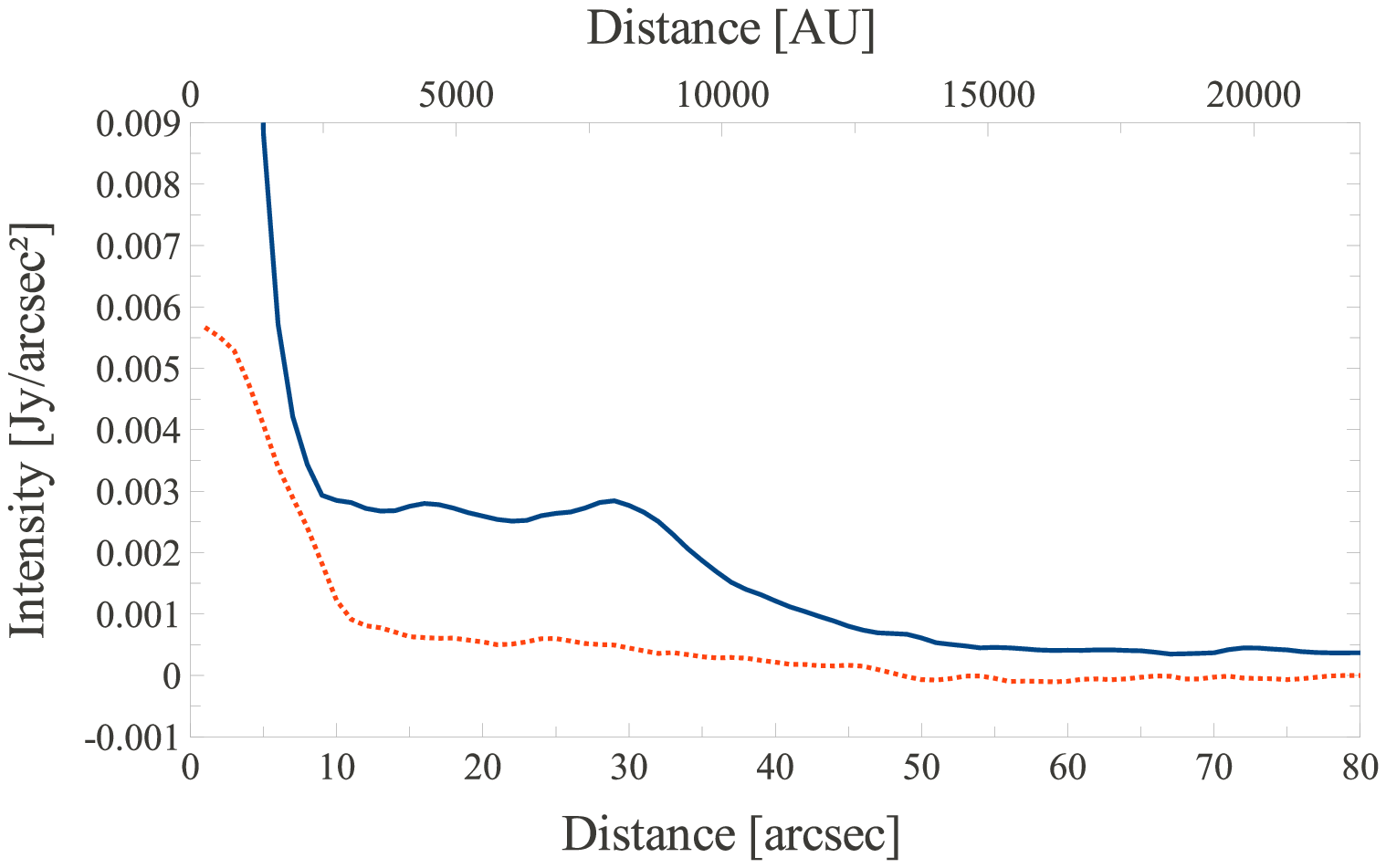}
     \caption{\label{Fig:txpsc_xher_intensity}
Intensity plots for X~Her and TX~Psc. Upper plot: integrated intensity of X~Her in the direction of the northern blob (PA: $-7^{\circ}$ to $-17^{\circ}$) from the image of Fig.~\ref{Fig:xher_red_blue}. Lower plot: integrated intensity for TX~Psc
in direction of the blob/proper motion (PA: 228$^{\circ}$ to 250$^{\circ}$), from the image of Fig.~\ref{Fig:txpsc_red_blue}.  In both plots the (blue) continuous 
curve corresponds to 70 $\mu$m, the (red) dashed curve to 160~$\mu$m. The distances used to convert the angular scales into distances in AU are listed in Table~\ref{pmdata}.
}
\end{figure}

As noticed at the end of Sect.~\ref{Sect:blob}, an upstream blob is clearly visible in the Mach-10 isothermal-shock simulations of \citet{1998NewA....3..571B} (Fig.~\ref{Fig:Blondin}), as well as in several of the  simulations of \citet{1998A&A...338..273C}. Such a blob is present in the nebulae associated with both X~Her and TX~Psc. To investigate the nature of the emission associated with this blob, we performed aperture photometry in the 70 and 160~$\mu$m bands, for TX~Psc, where it is cleary  detached from the surrounding material
(lower right panel of Fig.~\ref{Fig:txpsc_scan}).
A radius of 5 arcsec was adopted to fully cover 
the blob. The background flux was estimated, from the periphery of the image,  to be $3\times10^{-4}$~Jy per pixel at 70~$\mu$m and $5\times10^{-4}$~Jy per pixel at 160~$\mu$m. With 
these values,  we get total fluxes of  
0.191 Jy at 70$\mu$m and 0.039~Jy at 160 $\mu$m, or a ratio $F_{\nu,70}/F_{\nu,160} = 4.89$ (see also the upper panel of Fig.~\ref{Fig:txpsc_xher_intensity}, which compares the intensity profiles in the 70 and 160 $\mu$m bands).
The observed light in this blob could in principle come from emission lines  (such as [O I] 63~$\mu$m, [O I]~145~$\mu$m, and [C II] 158~$\mu$m) excited by the shock\footnote{Remember that H$_\alpha$ emission as well as UV radiation (caused by excited H$_2$ molecules) are sometimes associated with AGB wind -- ISM interaction, as in the case of Mira \citep{2007Natur.448..780M} and R Hya \citep{2008ApJ...687L..33U, 2008PASJ...60S.407U}}. However, Spitzer/IRS spectroscopic observations of the R~Hya bow shock did not detect atomic lines \citep{Ueta_galagb}, and the detected far-IR radiation from the bow suggests instead the presence of very cold (20-40~K) material \citep{2009ASPC..418..463U}. The same conclusion was reached by \citet{1988ApJ...329L..93V} and \citet{1996ApJ...461..927D} in the context of OB runaway stars; they note that the dust trapped in the bow shock shell reradiates efficiently about 1\% of the star's bolometric luminosity at far-infrared wavelengths.

\begin{figure}[t]
\centering
   \includegraphics[width=8cm]{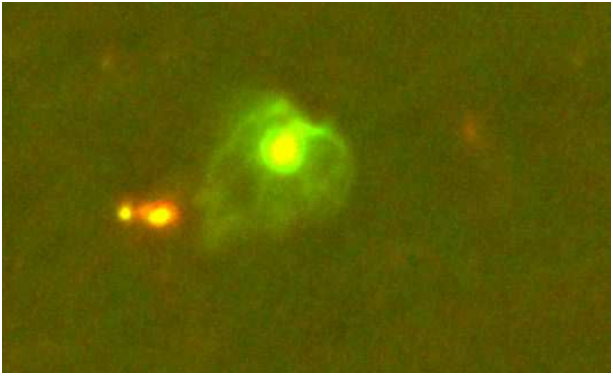}
     \caption{\label{Fig:xher_composite}
Composite image merging the PACS 70 and 160~$\mu$m images of X~Her (respectively in the green and red channels).
}
\end{figure}

Therefore, we make here the working hypothesis that the light emitted by the blob is from dust emission\footnote{Solving this issue would require the acquisition of a spectrum for this blob.}, and derive the corresponding dust temperature. From Kirchhoff's law, we may write the  dust emissivity $j_{\nu}$ as 
\begin{equation}
j_{\nu} = k_{\nu} \; j_{BB,\nu}(T),
\end{equation}
where $j_{BB,\nu}(T)$ is the black body emissivity at frequency $\nu$ and $k_{\nu}$ is the dust absorbing coefficient.
Since TX~Psc is a carbon star, we adopted the absorption 
coefficient for amorphous carbon as given by \citet{1981ApJ...245..880D}, who provides the $Q$ factors (ratio of the absorption cross section to the geometrical cross section) as a function of wavelength. Since $k_{\nu}$ is proportional to $Q$ and we are only interested in ratios, we may write
\begin{equation}
\frac{F_{\nu,70}}{F_{\nu,160}} = \frac{Q_{70}\; j_{BB,\nu (70)}(T)}{Q_{160}\; j_{BB,\nu(160)}(T)},
\end{equation}
where $Q_{70}/Q_{160} = 512/219 = 2.34$. Thus the dust temperature is that satisfying the relation 
\begin{equation}
\frac{ j_{BB,\nu (70)}(T)}{j_{BB,\nu(160)}(T)} = 4.89/2.34 = 2.09.
\end{equation}
A temperature of 81~K for 
the blob satisfies the above condition. Since the absorption coefficient is for dust containing only 
amorphous carbon, the temperature derived above must be seen as a rough estimate. 
The same analysis for X~Her, adopting the ratio $Q_{70}/Q_{160} = 185/25=7.4$ for astronomical silicates \citep{1985ApJS...57..587D}, yields a temperature of 40~K.
The composite colour image of X~Her presented in Fig.~\ref{Fig:xher_composite}
does not show evidence of any large temperature variations in the nebula, which is dominated by the emission at 70~$\mu$m.

\subsection{Wilkin fitting of the bow shock interface}
\label{wilkin}

The far-IR parabolic-like arc structures around X~Her and TX~Psc can be
interpreted as the interface regions between the 
ISM and AGB winds, as already recognized for R~Hya \citep{2006ApJ...648L..39U}, 
R~Cas \citep{2010A&A...514A..16U}, and IRC~$+$10 216 \citep{2010A&A...518L.141L}.
Assuming that (1) the ram-pressure balance is established between the 
ambient ISM and AGB winds and (2) momentum is conserved across a
physically thin shock interface (i.e.\ the shock is radiatively
cooled efficiently), one can express the bow shock shape analytically as
a function of the latitudinal angle $\theta$ measured from the apex of
the bow with respect to the position of the central star as formulated
by \citet{1996ApJ...459L..31W}
\begin{eqnarray}\label{bow}
 R(\theta) = R_0 \frac{\sqrt{3(1-\theta\cot\theta})}{\sin\theta},
\end{eqnarray} 
where $R_0$ is the {\sl stand-off distance} between the star and bow
apex, and  defined as 
\begin{eqnarray}\label{standoff}
 R_0 =\sqrt{\frac{\dot{M}v_{\rm w}}{4\pi\rho_{\rm ISM} v_{*,{\rm ISM}}^2}}
\end{eqnarray}
for which $\dot{M}$ is the mass-loss rate,
$v_{\rm w}$ is the isotropic stellar wind velocity, 
$\rho_{\rm ISM}$ is the ambient ISM mass density, and 
$v_{*,{\rm ISM}}$ is the relative velocity of the star 
{\sl with respect to the ambient ISM}.

Because the observed surface brightness enhancement represents
approximately a conic
section of an inclined bow shock paraboloid by the plane of the sky
intersecting the central star,\footnote{Assuming the observed
surface brightness is mainly due to thermal dust emission, the surface
brightness is roughly proportional to the dust column density along the
line of sight in the optically thin limit, which is true for the far-IR
wavelengths. 
If the dust distribution is an inclined paraboloid, the column
density becomes the greatest at the intersection of the structure with
the plane of the sky and therefore the observed surface brightness
distribution represents the conic section of the paraboloid.}
the inclination angle and {\sl deprojected} stand-off distance of the
bow shock paraboloid can (in principle) be determined by fitting the apparent shape of
the bow shock using the analytical Wilkin function (the detailed
description of the Wilkin fitting procedures may be found in Cox et al., in preparation).

As discussed in Sect.~\ref{Sect:blob}, the {\sl Herschel} images of the targets at higher spatial resolution
have resolved structures in the bow shock interface that resemble
vortices caused by instabilities at the contact discontinuity
\citep[e.g.,][]{2007ApJ...660L.129W}. 
In particular, the X~Her image shows that the bow shock interface is
bent in the direction of the star (Fig.~\ref{Fig:txpsc_xher_deconv}).
This bent-interface structure appears typical of bow shocks caused by
a large relative velocity between the star and the ambient ISM, 
$v_{*,{\rm ISM}}$, as shown in hydrodynamical simulations 
by \citet{2007ApJ...660L.129W} (Fig.~\ref{Fig:Wareing}).

The presence of these bent structures obviously makes the application of
the Wilkin fitting problematic.
Nevertheless, we performed the Wilkin fitting of the bow shock with
and without the bent region and obtained essentially the same results.
While we present the results of Wilkin fitting below assuming that 
a conic section of the bow shock paraboloid is still recoverable from
the overall shape of the observed far-IR surface brightness
distributions, we recall that the observed bow shocks of X~Her and
TX~Psc show asymmetric structures that were not seen in any previous AGB
wind-ISM interaction cases.

Using the method outlined by Cox et al. (in preparation), the Wilkin fitting then yields a best-fit bow shock
paraboloid with $R_0 = 36\arcsec$ (corresponding to 4930~AU at the
parallactic distance of 137~pc) at the position angle (PA; defined as degrees E of N) of $330^\circ$ for X~Her and 
with $R_0 = 37\arcsec$ (corresponding to 10175~AU at the parallactic distance of 275 pc) and 
at the PA of $238^\circ$ for TX~Psc.

The PA obtained above (characterizing the relative motion
of the AGB star with respect to the ambient ISM as determined from the 
apparent {\sl heliocentric} orientation of the bow shock paraboloid, and
denoted $v_{*,{\rm ISM}}$) must be compared 
with the PA  of the stellar  space motion in the LSR ($v_{*,{\rm LSR}}$), as listed in Table~\ref{pmdata}.
They are, for both stars, roughly consistent with each other: 
330$^\circ$ and 309$^\circ$ for X~Her, and 238$^\circ$ and 243$^\circ$ for TX~Psc.
The 20$^\circ$ discrepancy observed for X~Her may be an indication that the ISM around that star is not at rest in the LSR; however, we do not attempt to infer the velocity of that ISM flow, because of the uncertainties associated with Wilkin fitting for
such peculiar shock shapes as observed around the two stars under consideration.

Assuming as a first approximation that the ISM is at rest in the LSR, 
(thus $v_{*,{\rm ISM}} = v_{*,{\rm LSR}}$, the latter value being listed in Table~\ref{pmdata}), Eq.~\ref{standoff} may then be used to derive the ISM density close to X~Her and TX~Psc, knowing their mass loss rate and wind velocity.
Adopting for these values
$\dot{M} = 1.5 \times 10^{-7}$ M$_{\odot}$ yr$^{-1}$
and $6.5$ km s$^{-1}$ for X~Her \citep{2003A&A...411..123G} and 
$\dot{M} = 9.1 \times 10^{-8}$ M$_{\odot}$ yr$^{-1}$
and $10.5$ km s$^{-1}$ for TX~Psc \citep{1993ApJS...87..267O}, we find $n_H = 0.67$ and 0.16~atoms~cm$^{-3}$ for X~Her and TX~Psc, respectively. These values appear quite reasonable, given the high Galactic latitudes of these two stars (Table~\ref{pmdata}).


\section{Bipolar outflows and global picture}
\label{Sect:bipolar}

\subsection{X Her}
\label{Sect:XHer}

An extensive description of the circumstellar environment of X~Her 
can be found in \citet{2006MNRAS.365..245G} and \citet{Matthews2011}, 
and only the most noteworthy features will be repeated here. 
X~Her has distinctly double-component lines in CO ($J$\,=\,$2-1$) and 
($J$\,=\,$3-2$), i.e., a narrow feature centred on a broader feature, 
with $v_w \sim 3$ and $\sim9$~km~s$^{-1}$, respectively \citep[][] 
{1996A&A...310..952K, 1998ApJS..117..209K, 1999A&AS..138..299K, 
2002A&A...391.1053O, 2003A&A...411..123G, 2005ApJ...620..943N,
Castro-Carrizo2010}. \citet{2003A&A...411..123G} observe a similarly double-component SiO($v=0, J = 2-1$) line; from CO data, they derive mass-loss rates of $1.5\times10^{-7}$~M$_\odot$~y$^{-1}$ and 
$0.4\times10^{-7}$~M$_\odot$~y$^{-1}$, respectively, for the broad 
and narrow components. 
Although \citet{1998ApJS..117..209K} argued that this complex line 
profile may be caused by episodic mass loss with highly varying gas 
expansion velocities, it seems more likely now that it results from a 
complicated flow geometry. 

\begin{figure}
\centering
\includegraphics[width=9cm]{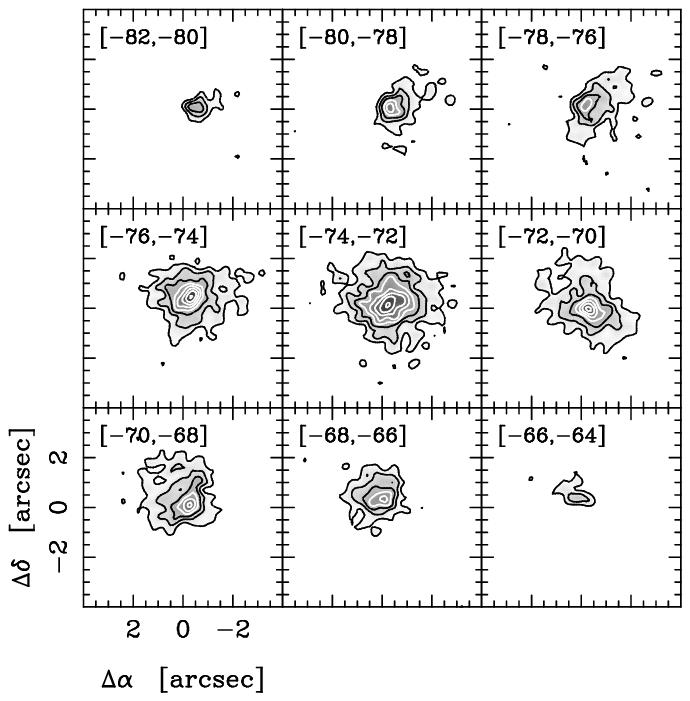}
\includegraphics[width=9cm]{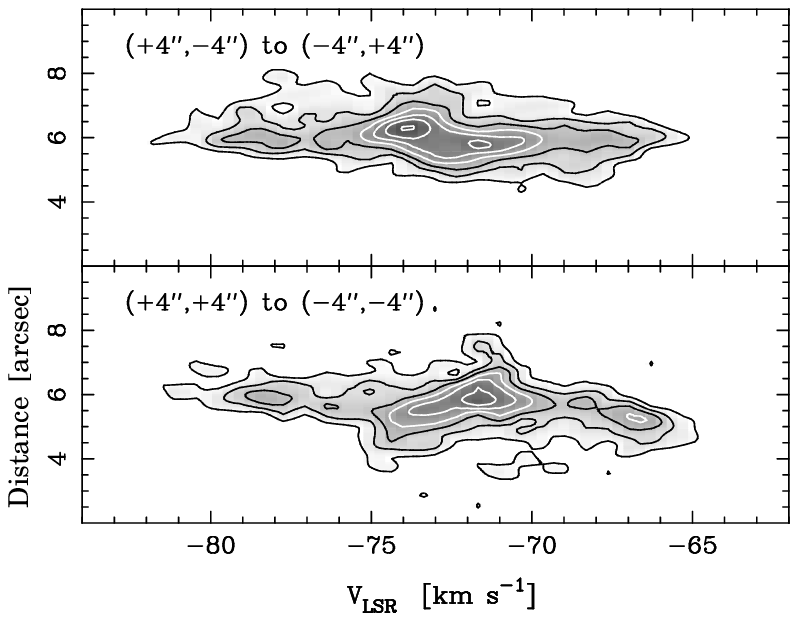}
     \caption{\label{Fig:XHerjet3}
Maps from the Very Large Array in the SiO $(v=0, J=1-0$) line for 
X~Her. The upper panels show 2 km\,s$^{-1}$ wide channel maps, and the 
lower panels show position (in arcsecs, along the vertical axis) -- 
velocity (in km/s, along the horizontal axis) diagrams along the NE -- 
SW [labelled (+4,+4) to (-4, -4)] and NW - SE [labelled (+4,-4) to 
(-4, +4)] directions. Velocities are with respect to the local 
standard of rest (systemic velocity of X~Her is -72.5 km/s). The lower 
panels show evidence of a bipolar outflow in the NE -- SW direction 
(thus perpendicular to the space motion) and of a rotating disk in the 
NW -- SE direction. The contour 
increments are 10~mJy~km~s$^{-1}$ per beam for the channel maps in the upper panel and 5~mJy per beam for 
the position-velocity diagrams in the lower panel. The beam size is 0.45$\arcsec \times 0.40\arcsec$ (with a position angle of $-88\deg$). 
}
\end{figure}

The spatial information provided by the CO radio line observations of 
\citet{1996A&A...310..952K} suggests that the broad plateau is a 
bipolar outflow oriented in the SW -- NE direction, with the SW side 
facing towards us, a structure that was confirmed by 
\citet{2005ApJ...620..943N}. The most recent CO radio line interferometry data of 
\citet{Castro-Carrizo2010} clearly resolves an hour-glass structure in 
the higher-velocity emission with a position angle of $45^\circ$, 
i.e., this indicates that a collimated bipolar wind in this direction 
has shaped the original AGB circumstellar envelope. The inclination 
angle of the outflow is likely to be small. \citet{Castro-Carrizo2010} 
find no evidence of a (rotating) disk perpendicular to the outflow 
axis. We  obtained SiO $(v=0, J=1-0)$ line data towards X~Her 
using the Very Large Array (details about these data and their reduction are given in the Appendix). 
These data probe a region smaller than 
that of the CO data because the SiO molecules are confined to a 
smaller region and they are effectively excited only in the inner more 
dense parts of the circumstellar envelope. Nevertheless, the channel 
map data presented in Fig.~\ref{Fig:XHerjet3} show a structure very 
similar to that of the CO data, a bipolar outflow at position angle $\approx\,45^\circ$ and a hint of an hour-glass structure. 
The position-velocity diagram along the outflow axis is also very similar to that 
obtained from the CO line data (Fig.~\ref{Fig:XHerjet3}). 
Interestingly, the position-velocity diagram perpendicular to this 
axis shows a structure that could be interpreted in terms of a 
rotating disk. 

\citet{2006MNRAS.365..245G} and \citet{Matthews2011} detected the H~I emission at 21~cm from the circumstellar shell around X~Her, and found it to be  extended  along the direction of the space motion ($\ge 6\arcmin$ or about 0.24~pc). \citet{Matthews2011} found, in their H~I data and on an arcmin scale,  some evidence for the continuation of the bipolar outflow observed on small scales ($\pm 10\arcsec$) in the molecular CO and SiO lines.

The surprising picture that emerges from the combination of all the data, including the new {\sl Herschel} data set, is that of a  bipolar outflow in the innermost region ($\pm 10\arcsec$), that appears to be almost exactly perpendicular to the far-IR and H~I nebula oriented along the direction of space motion and caused by the interaction between the stellar wind and the ISM. 
It is not at all clear whether this is just a chance coincidence, or whether there could be a causal relationship between the orientation of the bipolar outflow and the space motion. The absence of  any sign whatsoever of  the bipolar outflow visible in the {\sl Herschel} images would tend to dismiss the latter possibility. 
The existence of a bipolar outflow seen in CO and SiO molecular lines raises another question: how can this outflow be reconciled with the bow shock seen all around the star and signaling that mass loss is (was?) roughly isotropic? Two solutions may be envisioned: (i) even in the presence of a bipolar outflow, mass continues to be ejected in all directions, albeit probably at a lower rate; (ii) the bipolar outflow has formed more recently than the features seen in the far IR.

\subsection{TX Psc}

The close environment of TX~Psc, as probed by the molecular lines, appears to be quite complex. The observed line profiles of the CO line emission towards TX~Psc are indeed peculiar, as first noted by \citet{1989A&A...218L...5H}. 
As a consequence, the expansion velocity derived from the observed radio line profiles range 
from 7.5 to 12.2 km~s$^{-1}$ (see Table~2 of \citeauthor{2001A&A...368..969S} \citeyear{2001A&A...368..969S}, and \citeauthor{1993A&AS...99..291L} \citeyear{1993A&AS...99..291L}) depending on the transition and the telescope used. 
\citet{1989A&A...218L...5H} mapped the circumstellar CO line 
emission around this object and interpreted the observed 
asymmetries as being produced by either a bipolar outflow 
or a highly clumped wind. No detailed modelling exists for 
the circumstellar shell around TX~Psc, and its evolutionary status remains uncertain. 

The asymmetry detected by \citet{1989A&A...218L...5H} and possibly attributable to a bipolar outflow  is
oriented along the NE -- SW direction, with the red- and blue-shifted part of the emission separated by about 10 arcsec. However, there is some ambiguity about this direction in the original paper. Interestingly enough, both lunar occultation data \citep{1995A&A...301..439R}  and Come-On+ adaptive-optics observations \citep{1998A&A...338..132C} claim as well that the source is asymmetric. The September 1994 Come-On+ observations in the $K$ band clearly reveal  a secondary source  0.35$\arcsec$ to the SW (with a position angle of $241^\circ$) of the primary source, with a flux ratio of 2\%. 
The lunar occultation observations indicate that the brightness profile of TX~Psc has a structure that becomes increasingly more complex when going from the $V$ band to the $L$ band, where it is double peaked (on an angular scale of a few milliarcsecs). The $V$ band reveals some secondary peaks on either side of the central core. In the visual band, \citet{1996AJ....111..936H} and \citet{1999AJ....117.1890M} could not, however, resolve TX~Psc between 1993 and 1997 with a resolving limit of 0.054$\arcsec$. 

The gas mass-loss rate of TX~Psc has been derived by several authors. 
\citet{1993ApJS...87..267O} found a very moderate gas mass-loss rate of $9.1\times10^{-8}$~M$_\odot$~yr$^{-1}$ from CO observations and a wind expansion velocity of 10.5~km~s$^{-1}$. \citet{1993A&AS...99..291L} found $5.6\times10^{-7}$~M$_\odot$~yr$^{-1}$ and in a later paper \citet{2001A&A...368..969S} derived an expansion velocity of 7.5~km~s$^{-1}$, with a complex CO line shape that prevented them from attempting to derive the mass-loss rate. 

The new Herschel-PACS data described in Sect.~\ref{Sect:data_analysis} confirm the presence of an axisymmetry in the far IR with a NE -- SW axis and on an arcmin scale.  However, in contrast to the situation prevailing for X~Her, the asymmetries observed around TX~Psc on an arcsec scale have the same orientation as the proper motion and the far-IR nebula. Nevertheless, an almost circular  shell, with a radius  of about 17$\arcsec$, surrounds TX~Psc and does not bear any signature
of the asymmetries observed on a much smaller angular scale.

\section{Conclusions}

The PACS images of the O-rich star X~Her and the C-rich star TX~Psc obtained by \textit{Herschel} at 70 and 160~$\mu$m
reveal a complex structure with signatures of interaction between the AGB wind and the ISM. Their space velocities, corrected from the solar motion, are 90 and 67~km~s$^{-1}$, respectively.
Its velocity locates X~Her among the high-velocity group defined by \citet{2005A&A...430..165F} in their kinematic study of giant stars from the Hipparcos Catalogue.   
The most noteworthy feature is, for both stars, the 
presence of a clump located right along the direction of the space motion. It is very likely the signature of instabilities arising in the shock front 
between the ISM and the AGB wind. The kinematical age of these clumps is (at least) about 2000~yrs for both stars. Although emission lines excited by the shock may contribute to the observed infrared light, assuming instead that only dust emission is involved yields a dust temperature of about 40~K for X~Her and 80~K for TX~Psc. The definite identification of the physical process at the origin of the light emitted by the bright clumps must await their detailed spectroscopic analysis.  
Further downstream along the bow shock, the PACS images reveal the presence of fainter clumps that may be associated with Kelvin-Helmholtz vortices peeling off the bow shock and moving downstream. There is actually a striking similarity between the PACS  image of X~Her, and the predictions of the hydrodynamical simulation of \citet{2007MNRAS.382.1233W} for the case 
$v_{*,{\rm ISM}}=125$~km~s$^{-1}$ (as compared to the observed value of 90~km~s$^{-1}$),  $n_H=2$~cm$^{-3}$, and a mass-loss rate of $5\times10^{-6}$~M$_\odot$~yr$^{-1}$ (a factor of 30 larger than the \textit{current} mass loss rate, however), where a front shock is bent towards the mass-losing star along the upstream direction, and K-H vortices are downstream along the bow shock. The strong blob visible upstream  of TX~Psc may be an indication that the wind and the ISM flow are highly supersonic, according to the simulations of \citet{1998NewA....3..571B}.

The Wilkin fitting of the bow shock shape is made difficult by its disturbed, highly structured shape (K-H instabilities?) 
and its reverse bending centred on the upstream direction. 
Nonetheless, the position angle of the bow shock obtained by a Wilkin fitting is roughly consistent with the position angle of the space motion of the star with respect to the LSR (the largest discrepancy is 20$^\circ$ for X~Her), indicating that the ISM is not far from being at rest in the LSR local to the stars.

A very puzzling conclusion of the present analysis is  that for both stars the axisymmetry caused by the space motion and its interaction with the ISM seems to be maintained down to very small angular scales, even beyond the circular feature present on intermediate scales in both TX~Psc and X~Her:
\begin{itemize}
\item In X Her, a secondary source almost perfectly aligned with the space motion  was discovered by earlier COME-ON+ observations, despite being located well within the ring observed with PACS.
\item Around TX~Psc, a bipolar outflow extending over a few arcseconds has been reported by radio observations (with some ambiguity however, about its orientation, but seemingly parallel to the space motion). A bipolar 
outflow was also  reported for  X~Her, and for that star,  the space motion and the  outflow are clearly perpendicular to each other. 
\end{itemize}

The existence of those correlations between the axisymmetry present in the  inner and outer 
circumstellar regions of TX~Psc and X~Her (despite circular structures being present on intermediate scales, possibly associated with the termination shock) is the most puzzling conclusion from the present work, for which we do not currently have any convincing explanation.
Similarly, it is rather surprising that the far-IR nebula contains a nearly perfectly circular shell (either interior to or identical to the termination shock), indicative of  (past) isotropic mass loss, whereas there is at the same time an inner bipolar  outflow. Could this be an indication that the mass loss became  bipolar quite recently?

\begin{acknowledgements}
We acknowledge with thanks the variable star observations from the AAVSO International Database contributed by observers worldwide and used in this research. This work was supported in part  by the
Belgian Federal Science Policy Office via the PRODEX Programme of ESA (No. C90370). WN acknowledges funding by the FWF under project number P21988-N16. FK acknowledges funding by the Austrian Science Fund FWF under project number P23586-N16, RO under project number I163-N16, JH under project number P19503-N16 .
The Very Large Array is operated by the National Radio Astronomy
Observatory, which is a facility of the National Science Foundation
operated under cooperative agreement by Associated Universities, Inc.
\end{acknowledgements}

\bibliographystyle{aa}
\bibliography{biblio}

\appendix
\section{SiO line VLA observations of X Her}

X Her was observed in the rotational lines of SiO using the Very Large
Array in C configuration.  As part of a 10-hour observing run on 2001
August 24, about one hour was spent to observe the $(J= 1 - 0,
v=0)$ transition at 43.424~GHz in dual circular polarization.  A long
run of nine hours in right-hand circular polarization (RCP) was observed
on 2004 March 2004.  We combined these data with an additional hour
dual polarization observations taken on 2004 April 22.

All these observations used a total bandwidth of 6.25 MHz ($\sim$ 40
km\,s$^{-1}$) centred on a Doppler velocity of $-$71 km\,s$^{-1}$
with respect to the Local Standard of Rest.  The bandwidth was divided
in 64 spectral channels, corresponding to a channel separation of
$\sim$ 0.67 km\,s$^{-1}$ (97.7 kHz), for all three observation runs.
Also common to the observations was the fast switching mode to the
phase reference source J1549$+$5038 using a typical cycle of $\sim$40
and 100 seconds for calibration and target source.  Using 3C286 as
a flux calibrator, we measured fluxes for J1549$+$5038 of 0.52
($\pm$0.01) and 0.61 ($\pm$0.01) Jy for 2001 August and 2004 March,
respectively.  We set it to be 0.62~Jy for the 2004 April observations.

The data were reduced using standard editing, calibration, and imaging
procedures available in the Astronomical Image Processing System.  The
RCP data was combined with the dual polarization data to Stokes\,I,
assuming that the thermal radiation is unpolarized (i.e., Stokes\,V =
0).  The concatenated visibility data set was imaged into a cleaned
cube of $\sim$13\arcsec\ $\times$ 13\arcsec\ over $-$87 to $-$54
km\,s$^{-1}$ with an angular resolution of about 0.5\arcsec\ and a RMS
noise of typically 2.2~mJy per channel. The results of these observations are displayed in Fig.~\ref{Fig:XHerjet3}.

\end{document}